\documentclass[conference]{IEEEtran}

\usepackage{algorithm}
\usepackage[noend]{algpseudocode}
\usepackage{graphicx}
\usepackage{amsmath,amssymb}
\usepackage{url}
\usepackage{balance}
\usepackage{color}
\usepackage{tabto}
\usepackage{caption} 
\makeatletter
\def\BState{\State\hskip-\ALG@thistlm}
\makeatother

\begin{document}
	
\title{Defending against Intrusion of Malicious UAVs with Networked UAV Defense Swarms}

\author{\IEEEauthorblockN{Matthias R. Brust$^1$, Gr\'{e}goire Danoy$^2$, Pascal Bouvry$^1$,
Dren Gashi$^2$, Himadri Pathak$^2$,
Mike P. Gon\c{c}alves$^2$}
\IEEEauthorblockA{$^1$Interdisciplinary Centre for Security, Reliability and Trust (SnT), University of Luxembourg, Luxembourg\\
$^2$Faculty of Science, Technology and Communication (FSTC), University of Luxembourg, Luxembourg
}}

\maketitle
	
\begin{abstract}
Nowadays, companies such as Amazon, Alibaba, and even pizza chains are pushing forward to use drones, also called UAVs (Unmanned Aerial Vehicles), for service provision, such as package and food delivery. As governments intend to use these immense economic benefits that UAVs have to offer, urban planners are moving forward to incorporate so-called \emph{UAV flight zones} and \emph{UAV highways} in their smart city designs. However, the high-speed mobility and behavior dynamics of UAVs need to be monitored to detect and, subsequently, to deal with intruders, rogue drones, and UAVs with a malicious intent.

This paper proposes a UAV defense system for the purpose of intercepting and escorting a malicious UAV outside the flight zone. The proposed UAV defense system consists of a defense UAV swarm, which is capable to self-organize its defense formation in the event of intruder detection, and chase the malicious UAV as a networked swarm.

Modular design principles have been used for our fully localized approach. We developed an innovative auto-balanced clustering process to realize the intercept- and capture-formation. As it turned out, the resulting networked defense UAV swarm is resilient against communication losses.
Finally, a prototype UAV simulator has been implemented. Through extensive simulations, we show the feasibility and performance of our approach.

\end{abstract}
	
\section{Introduction}
\label{sec:introduction}

Governments, companies, third parties, or even individual citizens could permit UAV owners to use their designated air space during a given time and decide for how long, and so rent space and time to service providers using UAVs. These licensed flight zones (or \emph{UAV highways}) then can be used for food and package delivery. We will see a diverse range of users and clients of this new kind of resource usage, in particular, in smart cities. 

However, the deployment of a large number of UAVs as independent entities comes with risk and security assurances. Due to the high dynamics of the system that cannot be done manually, increasing importance has to be put on the innovation, research, and development of a UAV defense system that consists of a monitoring system and defense UAVs \cite{mitchell2014adaptive}.

Such defense UAVs (dUAVs) can autonomously and collaboratively act as a defense swarm to deal with intruders, rogue drones, and UAVs with malicious intent. Malicious UAVs (mUAVs) can be intercepted, captured and escorted out of the flight zone.

\begin{figure}
					\begin{minipage}{0.24\linewidth}
						\centering
						\includegraphics[width=16mm]{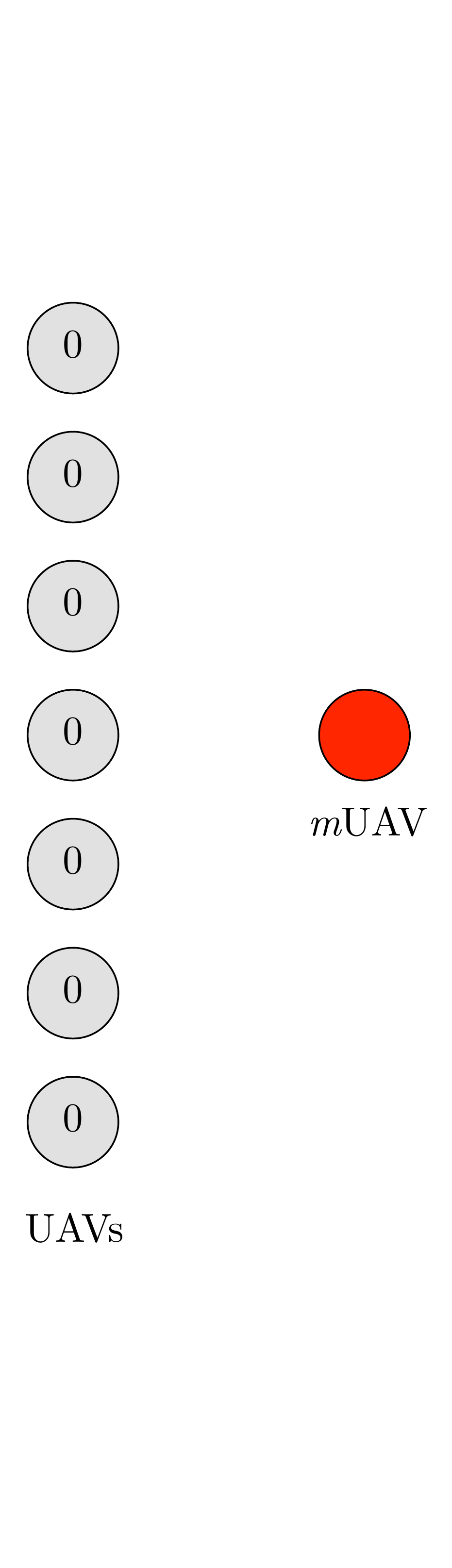} \\  
					\end{minipage}
					\vline
					\begin{minipage}{0.36\linewidth}
						\centering
						\includegraphics[width=31mm]{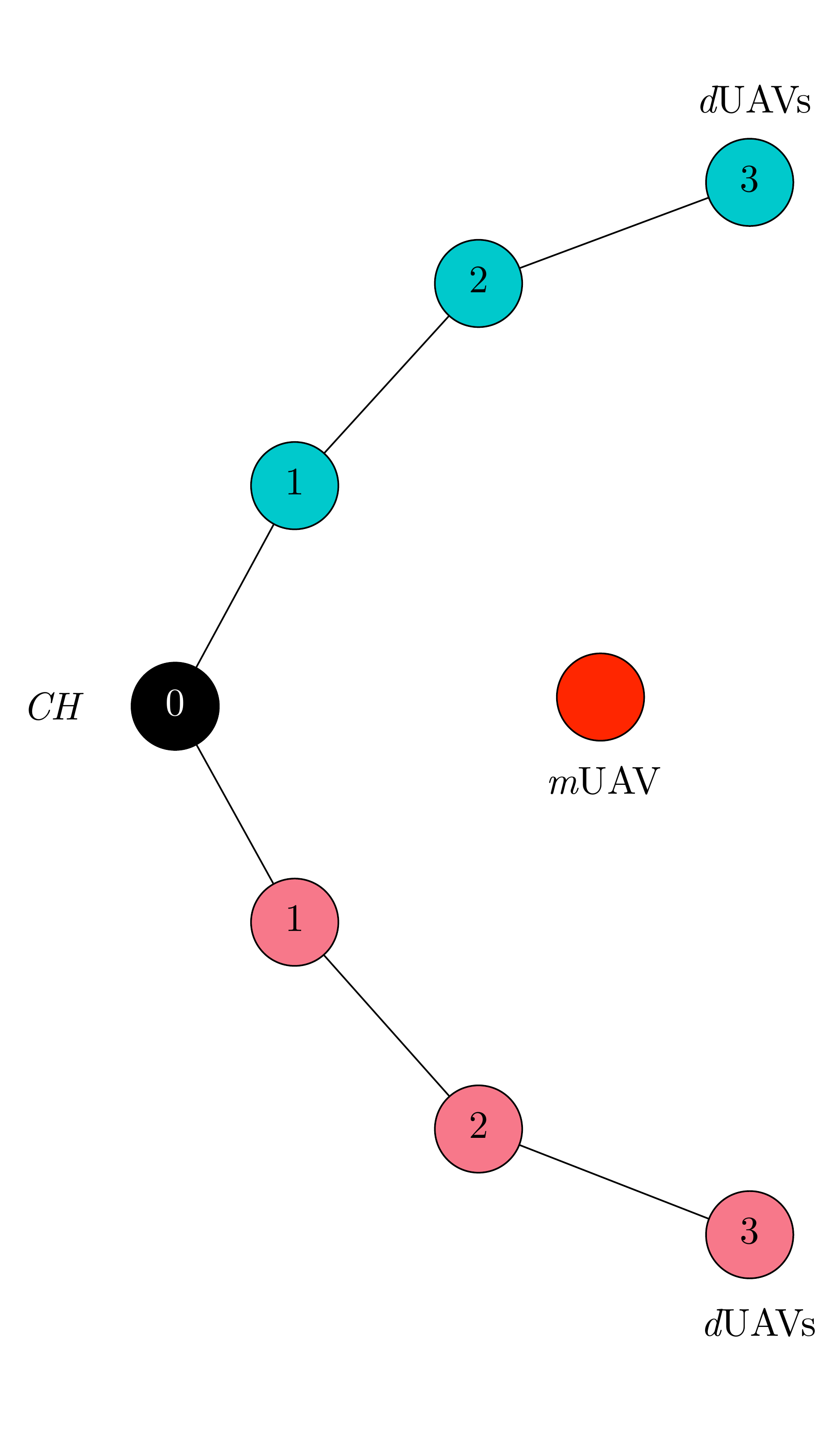} \\  
					\end{minipage}
					\vline
					\begin{minipage}{0.30\linewidth}
						\centering
						\includegraphics[width=31mm]{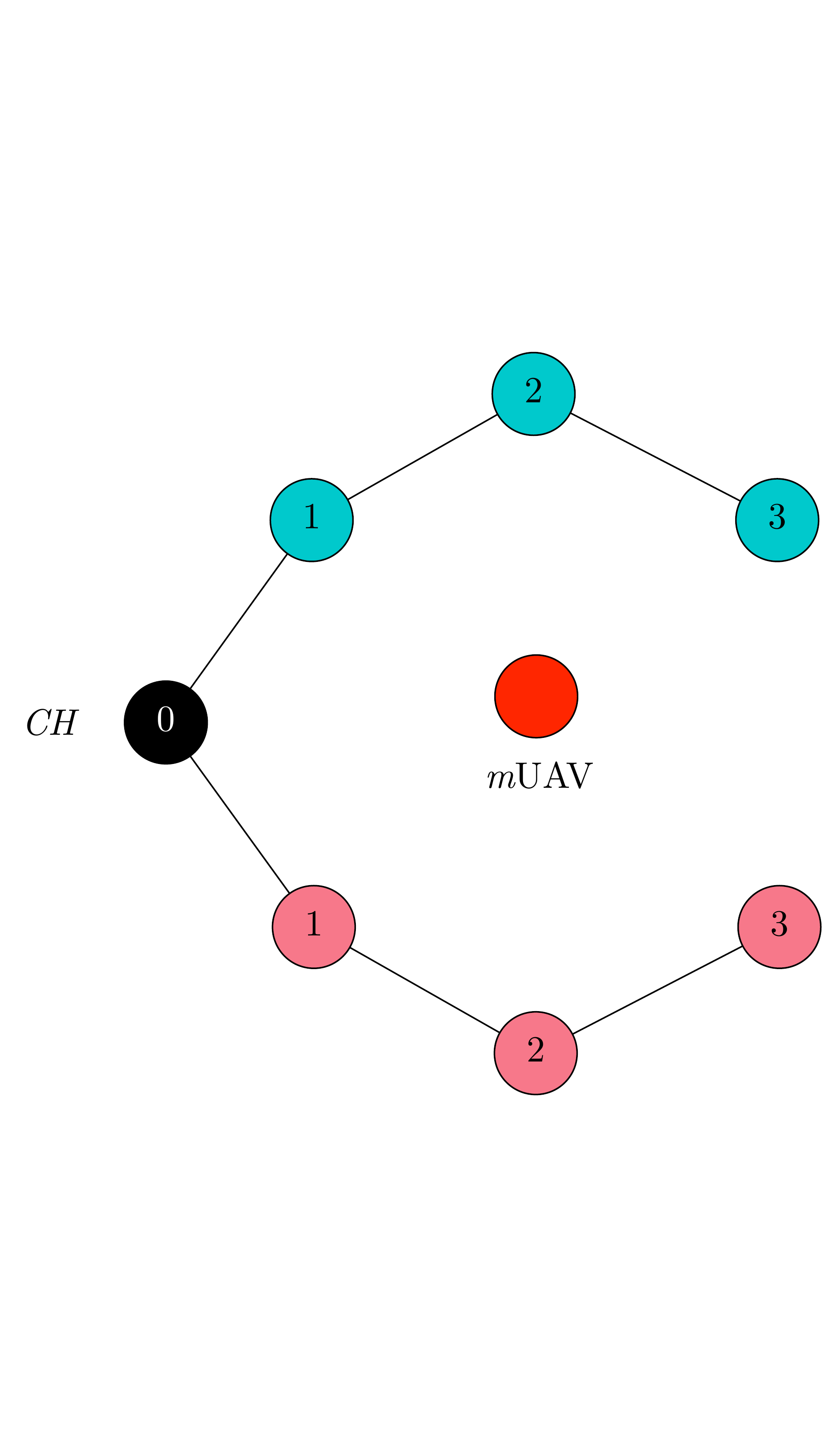} \\  
					\end{minipage}
					\caption{An illustration of the escort mission phases.}
					\label{fig:escort-mission-progress}
				\end{figure}

Our approach consists of a swarm of dUAVs that forms a three-dimensional cluster around the mUAV in a way to restrict its movement possibilities. Hereby, we assume that the mUAV is trying to avoid collision with dUAVs to maintain its functioning. By enclosing the mUAV, the movement possibilities are enforced by the dUAVs such that the mUAV surrounded by the dUAVs is moving outside the flight zone, thus escorting the mUAV (see Fig. \ref{fig:escort-mission-progress}).
		
A comprehensive UAV defense system is proposed, which is able to deploy auto-organized defense UAVs (dUAV) and create an intercept- and capture-formation to escort malicious UAVs (mUAV) outside the flight zone. 

The most outstanding features and contributions of the presented approach are the balanced clustering to realize the intercept- and capture-formation. Additionally, the approach consists of a modular design containing the phases such as deployment, clustering, formation, chase, and escort. All parts of the approach are fully localized, and the resulting networked defense UAV swarm is resilient against communication losses.

Our simulation results show that designated parameters have a greater influence over the total mission time that is needed to escort the malicious UAVs. 

The remainder of this paper is organized as follows. Section \ref{sec:related-work} reports on related work. In Section \ref{sec:definition}, we define the problem and describe the system model. Section \ref{sec:clustering}, \ref{sec:chase}, and \ref{sec:escort} explains each defense phase. In Section \ref{sec:prototype}, we describe our prototype and conduct a simulation study (cf. Section \ref{sec:sim-study}. We conclude with Section \ref{sec:conclusion}.

\section{Related Work}
\label{sec:related-work}
This section describes related work of UAV defense systems and UAV defense swarms. For the sake of completeness, we additionally include research on formation and positioning of UAVs.

The boom of UAV usage by private users and the forthcoming large-scale commercial applications bring new potential threats. UAV defense systems are therefore a hot topic when it comes to protect against intruders. Various approaches have already been considered to capture or neutralize a UAV, including net and laser guns, radio-wave jamming guns, to drones equipped with nets or even trained eagles \cite{Eagles}.

Such intruders can also include UAV swarms themselves. Since such enemy swarms are difficult to target and financially not worth being shot because a missile for a whole swarm is much more expensive than a drone, military forces are investing on swarm-on-swarm warfare tactics in order to attack other enemy swarms such as the U.S. Army \cite{10945/5700,10945/17462}.

Swarms of UAV designated for defensive tasks can be used for collapsing and trapping the enemy swarm. Collapsing is being done via communication jamming in order to disrupt the enemy swarm such that the individual drones get disintegrated and uncoordinated \cite{humphreys2015statement}. This can be done, because autonomous UAV systems essentially rely on two types of wireless links, which are the command
link to the operator and the navigation signal link \cite{wesson2013hacking}. Recent research discusses how a defense mechanism can work to disturb or manipulate these before-mentioned links \cite{kerns2014unmanned}.

Regarding swarm control, research is conducted with flocking and swarming approaches \cite{brust2015networked}. Some researchers use both terms interchangeably, while others distinguish between them. In \cite{olfati2006flocking}, an overview is given on flocking and swarming algorithms. While the dynamics of flocking do not follow a pre-defined goal and therefore do not impose higher controllability levels, swarming can provide explicit elements of controllability to coordinate the swarm to the execution of a task, e.g., maneuvering to a specific tree, while avoiding obstacles on the way. 
Important work has been done by Reynolds, who introduced the standard model for swarm flocking \cite{reynolds1987flocks}.

There are some node positioning approaches to position UAV in the 3D-space, which are related to our work. For example, Brust et al. \cite{Brust2016VBCAAV} propose VBCA, a virtual forces clustering algorithm, which imitates the VSEPR model from molecular geometry for the arrangement of UAVs in a clustered swarm. The UAV's position is determined by the distance and role of its neighboring UAVs. VBCA assigns the role of a cluster head to one UAV. This central UAV acts as a connector influencing the entire topology of the network geometry while individual UAVs are only affected by their direct neighbors. VBCA is maximizing volume coverage, while maintaining \emph{advanced connectivity} within the clustered UAV swarm. 

\section{Problem Definition and System Model}
\label{sec:definition}
This section defines the system model on which our proposed approach is based. It describes the notations, definitions and assumptions used throughout this paper.

\subsection{Problem definition}  

For this paper, we assume a malicious UAV (mUAV) has been detected in the flight zone and a number of defense UAVs (dUAVs) have been instructed to initiate the defense mission by intercepting, capturing and escorting the mUAV out of the flight zone. The mUAV tends to escape, while avoiding collision, when it detects the dUAVs. 

\subsection{Definitions and notations}
		\begin{table}[H]
			\def\arraystretch{1.2}
			\begin{tabular}{c | c}		
							\textbf{Acronyms and Notation}				&       \textbf{Definition}		\\ \hline \hline
							 $|A|$ &  Number of elements in the set $A$	\\ 
							 $B$ & Set of branches of a cluster head															\\ 
                             $\beta$ & Enclosing angle		\\
                            BM-A &  Basic message - Accept		\\ 
                            BM-D &  Basic message - Discard		\\ 
							 CH & Cluster head																			\\ 
                            CM &  Control message		\\ 
							 $\hat{d}$ & Normalization of vector $\vec{d}$ where $\hat{d} = \frac{\vec{d}}{||\vec{d}||}$ \\ 
							 dUAV & Defense UAV											\\ 
							 $\epsilon_d$ & Collision threshold of dUAV										\\ 
                             $\epsilon_m$ & Collision threshold of mUAV
                             \\
							 Flight zone & Restricted area/space											\\ 
							 mUAV & Malicious UAV																\\ 
							 $N$ & Set of UAVs in the neighborhood										\\ 
							 $n_B$ & Pre-defined number of branches of a CH							\\ 
                             $r_F$ & Formation radius\\
							 $\left\Vert \vec{v} \right\Vert$ &  Magnitude of $\vec{v}$		\\
							 $w_B$ & Balanced clustering weight												\\ 
							 $w_K$ & KHOPCA weight																	\\  \hline
			\end{tabular}
		\end{table}
		
		\subsection{Properties}
		\begin{itemize}
			\item Every dUAV can have at most one parent and a child.
			\item CHs have no parent and can have up to $n_B$ children.
			\item Every CH stores a $n_B$.
			\item A branch is a dUAV that has a CH as parent.
			\item The length of a branch $x$ is defined by the number of dUAVs in parent-child relations starting from $x$. The length of $x$ is denoted as $|x|$.
			\begin{itemize}
				\item Example: CH$\rightarrow a\rightarrow b \rightarrow c \rightarrow d$. Then, $|a| = 4$.
			\end{itemize}
			\item Every dUAV has a clustering weight $w_B$, which is initially set to 0.
            \item A leaf is the last dUAV in a branch, has the largest $w_B$ in the branch and has no child.
			\item $w_B$ of a dUAV is defined as its position on a branch.
			\begin{itemize}
				\item Example: $ CH \rightarrow a \rightarrow b \rightarrow c \rightarrow d$ with $w_B$ assignment $(CH,0), (a,1), (b,2), (c,3), (d,4)$
			\end{itemize}
			\item The difference of the lengths of any two branches cannot exceed 1.
		\end{itemize}
        
        \subsection{Assumptions}
		\begin{itemize}
        	\item The mUAV has already been detected by every dUAV.
            \item Possible communication jamming capabilities used for attack or defense are not part of our formation-based approach.
            \item The mUAV has slightly lower top speed than dUAVs in order to avoid static locking.
            \item Every UAV actively tries avoiding collisions with each other.
            \item All UAVs have transmission, distance, relative positioning and absolute position sensing capabilities.
            \item A high-quality UAV monitoring system is in place that is capable to detect and identify malicious UAVs.

		\end{itemize}
	
\subsection{Communication model}
        Every UAV is equipped with a network adapter that can be used to establish a communication channel between UAVs. The communication itself could be realized with infrastructure-less and self-configuring UAV Ad hoc Networks (UAANETs) \cite{maxa2017survey} that are a subset of the Mobile Ad hoc Network (MANET) paradigm. For the sake of simplicity, we assume that every UAV has the capability of periodically scanning the surroundings by using a \textit{circular} transmission range. Furthermore, we assume a reliable communication channel.
        
\subsection{UAV monitoring system}
We assume that there is a UAV monitoring system in place to detect and identify the approximate location of the malicious UAV in the restricted area. That is, we assume that in the presence of a positive event the UAV monitoring system triggers the UAV defense system, the dUAVs deployment and then, initiates the creation/generation of the UAV defense swarm.       

\section{Our Approach}
\label{sec:approach}
Our approach consists of a swarm of dUAVs to form a three-dimensional cluster around the mUAV in such a way that the mUAV just has a minimum set of movement possibilities. Hereby, we assume that the mUAV is trying to avoid collisions with dUAVs to maintain its functioning. By enclosing the mUAV, the movement possibilities are enforced by the dUAVs such that the mUAV surrounded by the dUAVs are moving outside the flight zone, thus escorting the mUAV (see Fig. \ref{fig:escort-mission-progress} (c)).

The proposed approach follows a modular design, implementing four phases to realize the escort maneuver (task, problem), which are (1) clustering phase, (2) formation phase, (3) chase phase, and (4) escort phase.

The clustering and formation algorithms are executed simultaneously together during the whole escort mission, whereas the transition between the chase and escort phases are decided by the CH depending on the following conditions:

\begin{enumerate}
\item Chase phase to escort phase: The distance between the CH and the mUAV is lower than $r_F$.
\item Escort phase to chase phase: The distance between the CH and the mUAV is higher than two times $r_F$.
\end{enumerate}

A detailed description of the individual phases is provided in the next sections.

\section{Clustering Phase}
\label{sec:clustering}
The clustering algorithm is based on the KHOPCA clustering algorithm \cite{brust2007adaptive, brust2010lswtc, brust2008dynamic} with the key difference that the structure of the cluster remains balanced. We use KHOPCA for three main reasons. Firstly, it provides a leader election algorithm that creates cluster heads, which is the entry point for our clustering algorithm. Secondly, KHOPCA does not require weights to be unique. Implementing a \textit{simple} leader election would require such an assumption. Lastly, KHOPCA has been proven to be suitable for highly-dynamic networks, including UAV swarms for surveillance \cite{Gregoire:2015:MultiLevel}.

The cluster structure consists of the CH being in the middle of the cluster, acting as a coordinator of the whole cluster and a set of branches that are around the CH.

The reason for maintaining a balanced structure is the formation. Our goal is to construct a clustering that is suitable for the desired formation that looks like a closed hemisphere where the CH tries to enclose its branches in order to catch a mUAV. Therefore, the branches should ideally have the same length to be balanced. We also introduced the notation of a branch since it simplifies the modeling of the formation by considering a sequence of inter-connected dUAVs rather than single ones. The weighting constraint is defined as follows:
		\begin{center}
			$\forall b_i, b_j \in B \; : \; |\; b_i.length - b_j.length \; | \le 1$.\\
		\end{center}
The weighting constraint states that the difference of the lengths of any two branches cannot exceed 1.
In section \ref{sec:cluster-rebalancing} we illustrate how the re-balancing of the cluster works. Re-balancing is required due to unexpected connection losses. In section \ref{sec:formation} we elaborate on the formation.

The clustering is done fully locally at each UAV. We can distinguish between the three different states that an UAV can be in: UAV, dUAV and CH. The difference between UAVs and dUAVs is that UAVs have no parent, hence are not in a cluster and are searching for a parent while dUAVs are cluster members that are being coordinated by the CH for performing the escort mission. Every other dUAV will adapt its weight according to its parent. The dUAVs with weight $w_{B_i}$ are exactly $w_{B_i}$ hops away from the CH. Note that the weight of the clustering is not the same as the weight that KHOPCA provides. We differentiate between $w_B$ and $w_K$, where $w_B$ is the weight of the balanced clustering algorithm and $w_K$ is the one from the KHOPCA algorithm. We run both KHOPCA and our balanced clustering algorithm simultaneously. In the following, we elaborate on the different states of UAVs.
		
\subsection{Behavior: UAV}
Initially every UAV is parent-less and scans the neighborhood for a parent. Every UAV does not accept children by default and is flying to the mean position of the neighborhood. This \textit{flocking} ensures that UAVs nearby will be gathered together so that we can achieve bigger and fewer clusters. The flocking is described in the algorithm \ref{algo:flocking} and the behavior of UAVs is described in the algorithm \ref{algo:behaviorUAV}.
		
		\begin{algorithm}
			\caption{\label{algo:flocking}Flocking algorithm}
			\begin{algorithmic}
				\State $sum \gets \vec{0}$
				\For{$n \in N$}	
				\State $sum \gets sum + n.pos$
				\EndFor
				\State \textbf{end for}
				\State  $\mu_{pos} \gets sum \, / \, |N|$
                \State moveTo($\hat{\mu}_{pos}$)
			\end{algorithmic}
		\end{algorithm}
		
		\begin{algorithm}
			\caption{\label{algo:behaviorUAV}Behavior of UAVs (non-CH and parent-less)}
			\begin{algorithmic}
				\State {DoFlocking($N$)}
				\State $parents \gets \{n \in N \; | \, u.accept\}$
				
				\If {$|parents| > 0$}
                \State $duav = $ apply criterion to select $p \in parents$
				\State request connection with $duav$
				\EndIf
				\State \textbf{end if}
			\end{algorithmic}
		\end{algorithm}
		The UAVs scan the neighborhood for possible parents that accept children. If there exists more than one, we should consider to apply a criterion to choose one from $|parents|$.
		Our criterion is the minimal distance from the requesting UAV to the parent. Therefore, we sort the possible parent dUAVs in ascending order of distance. This enables a short communication channel and hence fewer potential connection losses. However, other criteria could be applied as well.
		
\subsection{Behavior: CH}
		As soon a UAV is elected as CH by the KHOPCA algorithm, it starts accepting children. CHs stop accepting further children if $n_B$ is reached. CHs then inform the children that they can now start accepting an additional child. We distinguish between the following two message types:

		\begin{enumerate}
			\item \textbf{Basic Message}	\\
			These messages are sent from the CH to its branches in order to trigger them into accepting an additional child (BM-A) or to discard (BM-D) the current child. Discarding a child at $w_{B_r}$ of a branch $r$ leads to the discarding of $|r|-w_{B_r}+1$ dUAVs since the message will be passed recursively to all children.
			
			\item \textbf{Control Message} \\
			CMs are recursively send from dUAVs to the CH in order to notify about a new child. 
		\end{enumerate}
		
CMs, as shown in Algorithm \ref{algo:behaviorCHs}, have the purpose of knowing the length of the branches of a CH which is crucial for the balancing mechanism.
		
		\begin{algorithm}
			\caption{\label{algo:behaviorCHs}Behavior of CHs}
			\begin{algorithmic}
				\State $accept \gets true$
				\While {$accept$}
				\If {$|B| = n_B$}
				\State send BM-A to all children
				\State $accept \gets false$
				\EndIf
				\State \textbf{end if}
				\EndWhile
				\State \textbf{end while}
			\end{algorithmic}
		\end{algorithm}
		
The algorithm runs until $|B| = n_B$. Then, CHs only act upon message receipt.
		
Upon receipt of a CM, CHs send a BM-A message to all its branches that have the minimal length among all branches. Hence, the leaves can start accepting a new child. This is how the CH ensures balancing. The balancing is a key feature that distinguishes our algorithm from KHOPCA. This behavior upon message receipt is described in the algorithm \ref{algo:CH-CM-message}.
		
		\begin{algorithm}
			\caption{\label{algo:CH-CM-message}CH: upon receipt of a CM}
			\begin{algorithmic}
				\State $senderUAV \gets \textit{sender of CM}$
				\State $b_s \gets getBranch(senderUAV)$
				\If {$b_s$ not null}
				\State $b_s.length \gets |b_s| + 1$
				
				\If {$|B| = n_B$}
				\State $min$ $\gets$ $min$(\{$b$ $\in$ $B$ : $|b|$\}) 
				\For{$b \in B$}	
				\If {$|b| = min$}
				\State send BM-A to $b$
				\EndIf
				\State \textbf{end if}
				\EndFor 
				\State \textbf{end for}
				\EndIf
				\State \textbf{end if}
				\EndIf
				\State \textbf{end if}
			\end{algorithmic}
		\end{algorithm}

\subsection{Behavior: dUAV}
The dUAVs have a parent and hence are in a cluster. They wait for incoming messages from the CH and are ready for chasing and formation. The dUAVs that are leaves in a cluster might still accept children. 
Let $d$ be a dUAV that has accepted a UAV $c$ as a child. Then, the following steps are executed:
		
		\begin{itemize}
			\item $c$ will join the cluster, hence become a dUAV
			\item $c.w_B \gets d.w_B + 1$
			\item $d.accept = false$
			\item $d$ will send a CM to its parent
		\end{itemize}
		
Let $d$ receive a CM. $d$ will propagate the CM to his parent.

Let $d$ receive a BM-A message. If $d$ has a child, it will no longer accept children and propagate the BM-A message to his child. If $d$ is a leaf, it will start accepting a child.
		
Let $d$ receive a BM-D message. If $d$ has a child, it will propagate the message to his child and discard the connection. $d$ will take the state of a UAV, thus resetting its $w_K$, performing flocking and searching for a new parent.
				
\subsection{Cluster re-balancing}
\label{sec:cluster-rebalancing}
Due to connection losses, the cluster can lose its balance. Therefore, we implemented a self re-balancing mechanism that keeps the cluster balance according to the weighting constraint.
		
\begin{figure}
			\makebox[\linewidth]{\includegraphics[width=65mm]{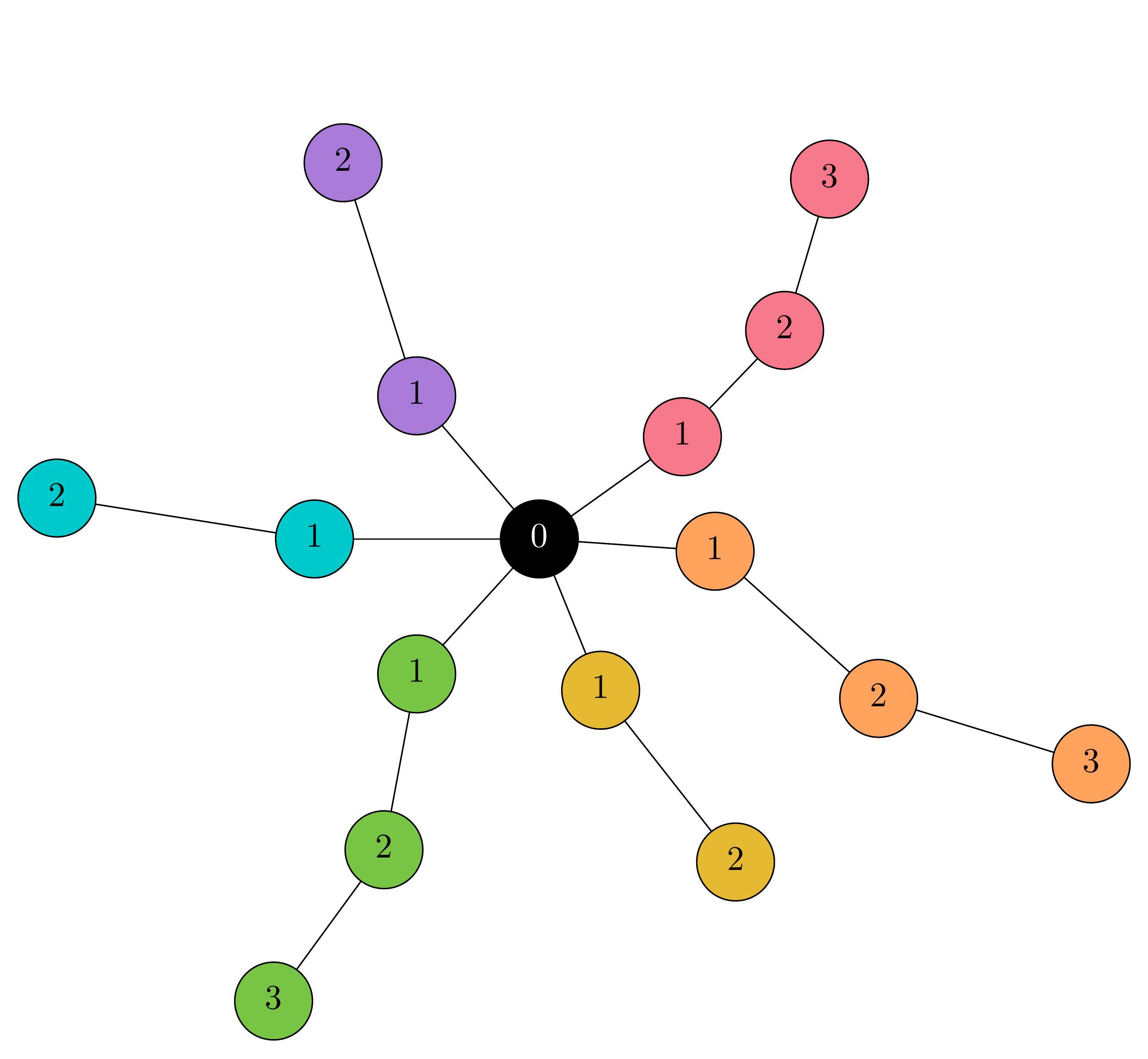}}
			\caption{Balanced cluster.}
			\label{fig:balancedCluster}
		\end{figure}

		In Fig. \ref{fig:balancedCluster} the re-balancing of a cluster is depicted.
        Algorithm \ref{algo:balancingCluster} shows the procedure of cluster re-balancing. Note that only CHs run the re-balancing algorithm.
		\begin{algorithm}
			\caption{\label{algo:balancingCluster}Cluster balancing}
			\begin{algorithmic}
				\State $min \gets min (\{b \in B : |b|$\})
				\For{$b \in B$}	
				\If{ $|b| > min + 1$ }
				\State $b.removeChildAt(min + 2)$
				\State $b.length = min + 1$
				\EndIf
				\State \textbf{end if}
				\EndFor
				\State \textbf{end for}
			\end{algorithmic}
		\end{algorithm}

\section{Formation Phase}
\label{sec:formation}
With the aim of escorting in mind, the swarm of UAVs chasing must ensure that the movement of the mUAV is restricted to one direction while the escort phase is under operation. 
To achieve the following, a formation model must be realized. This model should be resistant to any disruptions caused by the mUAV. Following from the assumption that $\epsilon_m>\epsilon_d$, one constraint we need to ensure is that the distance between the dUAVs within the formation is not too high which could allow the mUAV to escape.
 
\begin{figure}
		\begin{minipage}{0.40\linewidth}
			\centering
			\includegraphics[width=35mm]{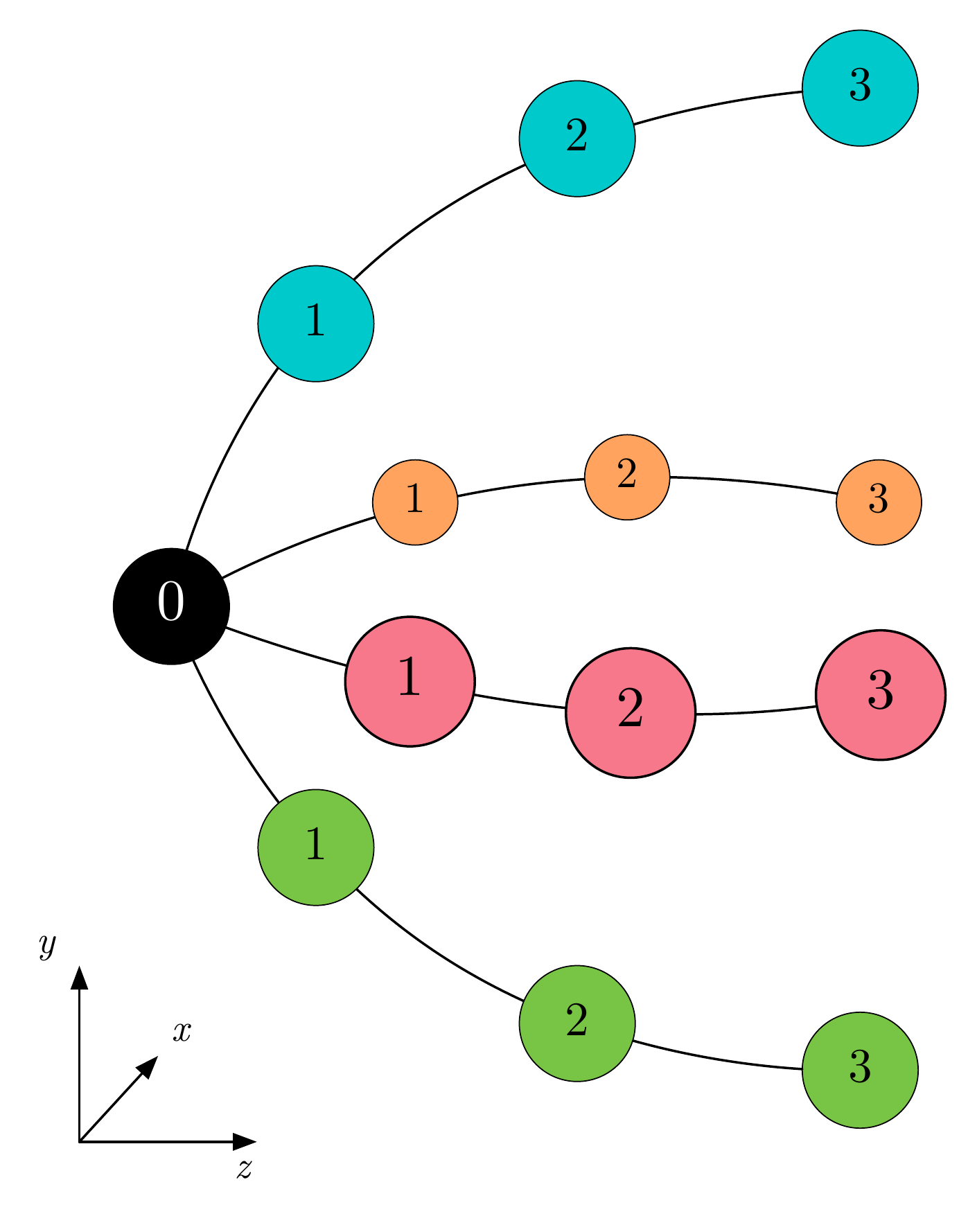} \\ (a)
		\end{minipage}
		\begin{minipage}{0.59\linewidth}
			\centering
			\includegraphics[width=50mm]{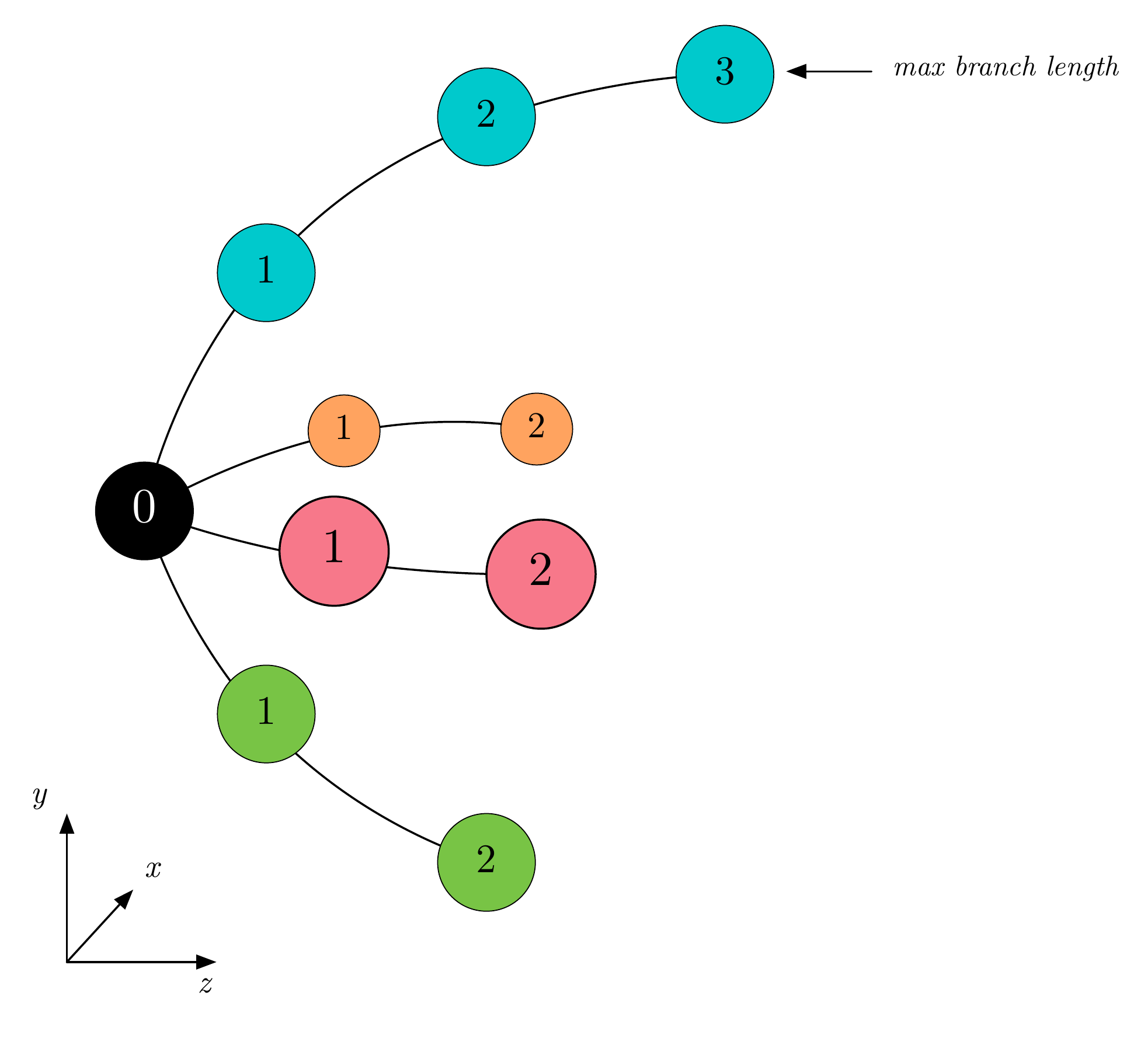} \\ (b)
		\end{minipage}
		\caption{Formation procedure.}
		\label{fig:formation}
	\end{figure}

The formation shape chosen for this particular problem is that of a hemisphere. 
The formation begins to take place while catching up to the mUAV and then proceeding to enclose when the cluster reaches a certain distance from the mUAV. This formation is depicted in Fig. \ref{fig:formation}a.

        \subsection{Calculating the cluster formation radius}
     Our goal is to place dUAVs on a branch equidistant to each other according to $\epsilon_d$ in order to minimize the escape directions of the mUAV. Firstly, we need to determine the maximum length of the branches to derive $r_F$. Let $max = max(\{b \in B : b.length\})$ be this number. 
     
     Suppose we inscribe a regular polygon into a circle.
      
     The branch does occupy only \(\frac{1}{4}^{th}\) of the imaginary circle. Every member of the branch lies on the edges of this circle. Therefore, if we were to mirror the singular branch in 2D along $y$-axis and then mirror the resultant along the $x$-axis, we would get a regular polygon with $n$ sides. Here, $n = 4 \cdot max$. Any regular polygon can be inscribed within a circle. 
          
     With this we can now find the $r_F$ with the following formula
 			\begin{center}
 				$r_F = \dfrac{\epsilon_d}{2 \sin(\frac{\pi}{4max})}$,
			\end{center}
where $r$ is the formation radius, $a$ is the length of a side in the polygon which is equal to the $\epsilon_d$.

      \subsection{Determining branch rotations}
\begin{figure}
			\begin{minipage}{0.40\linewidth}
				\centering
				\includegraphics[width=25mm]{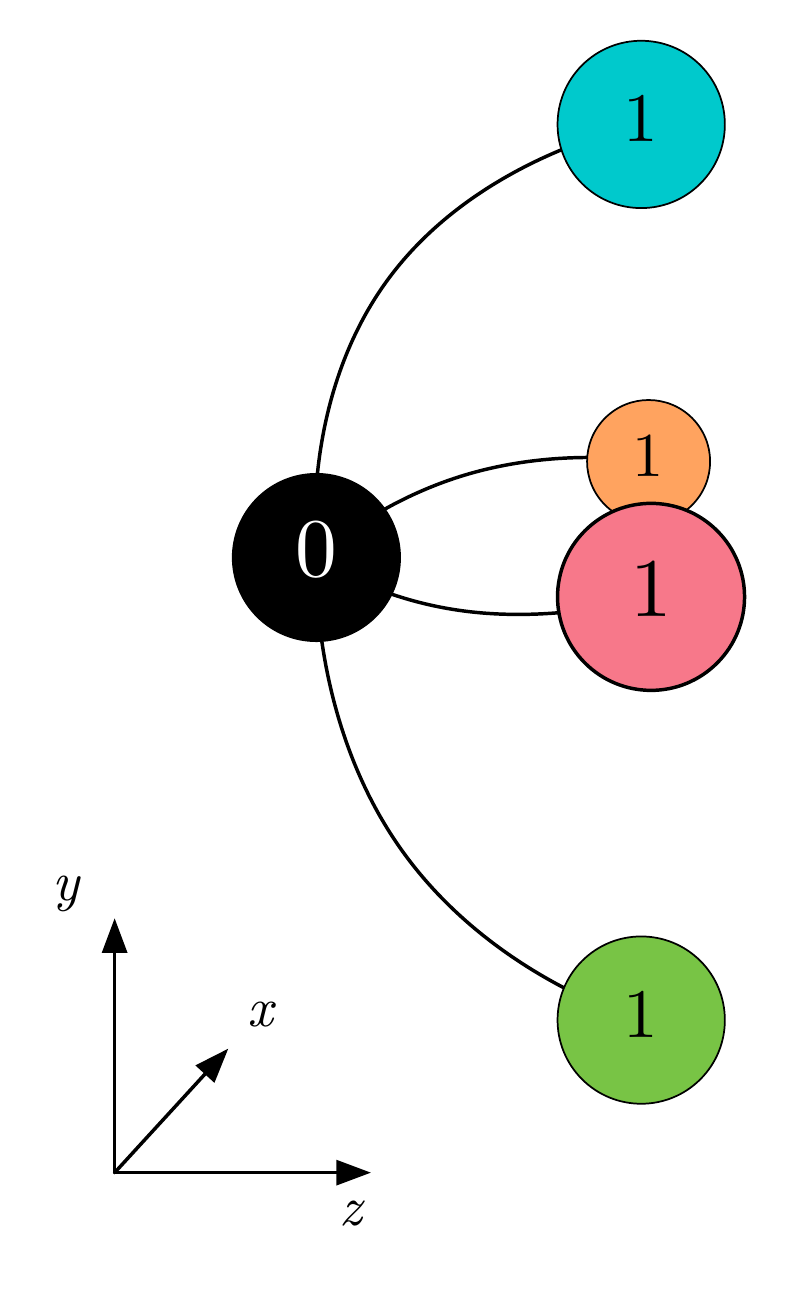} \\ Side view (a)
			\end{minipage}
			\begin{minipage}{0.59\linewidth}
				\centering
				\includegraphics[width=37mm]{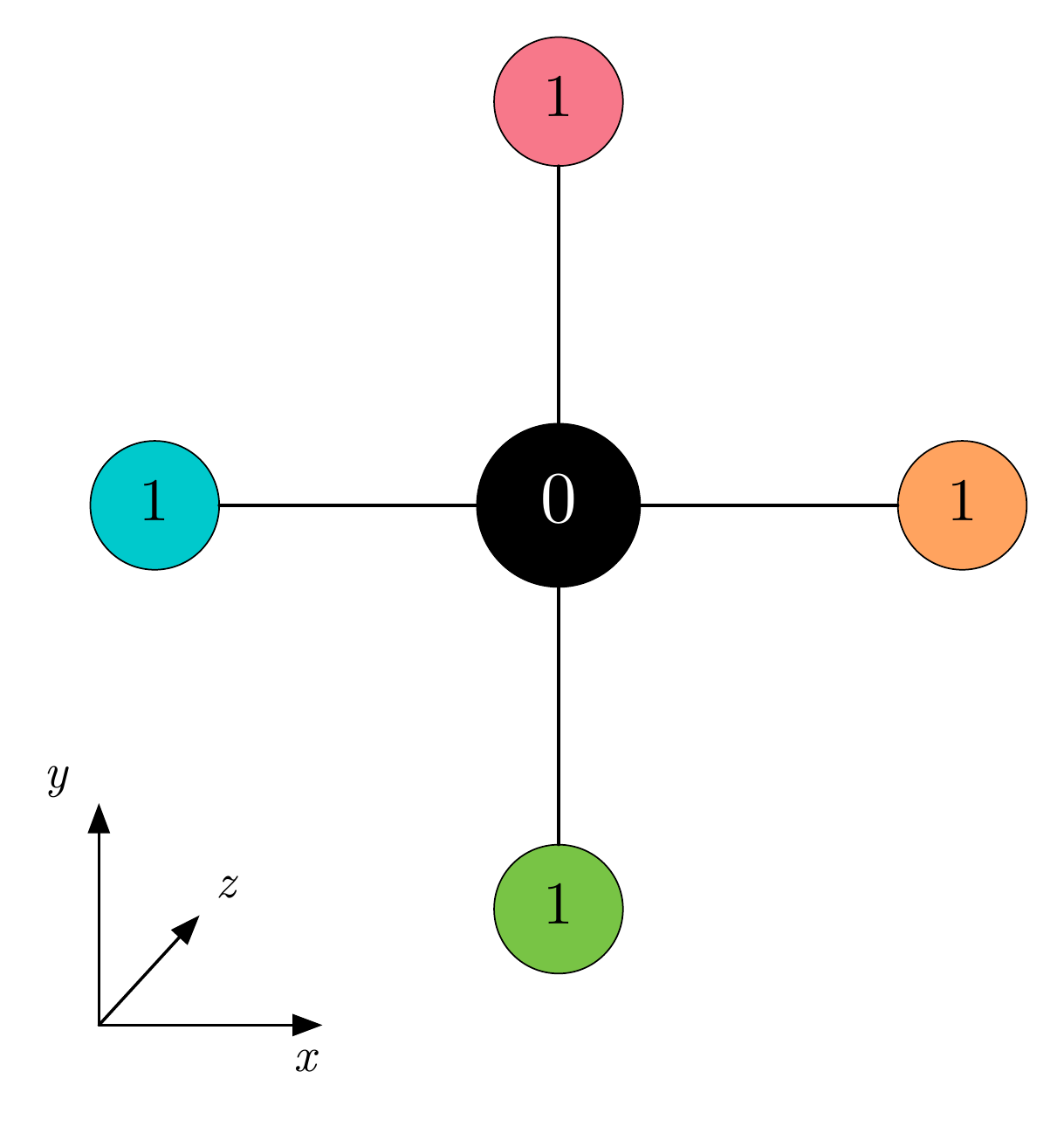} \\ Back view (b)
			\end{minipage}
			\caption{Side and back view of the branch rotation placement.}
			\label{fig:figure5}
		\end{figure}

\begin{figure}
			\makebox[\linewidth]{\includegraphics[width=60mm]{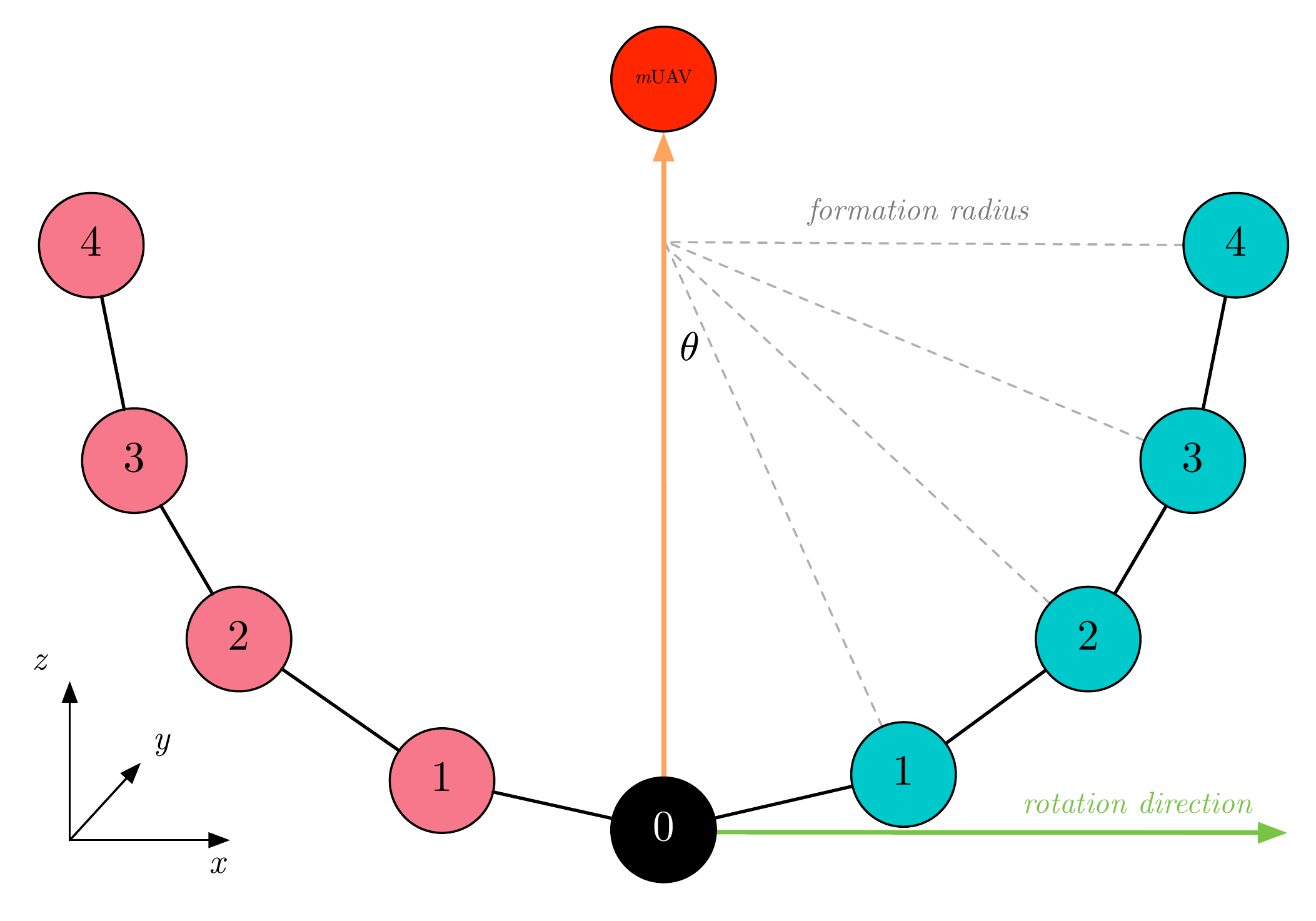}}
			\caption{Placing the children of the branches.}
			\label{fig:figure7}
		\end{figure}

Now that we have $r_F$, we know how far from the cluster head the branches are going to exceed. However to determine the positions of each branch relative to the cluster head along the $z$-axis as shown in Fig. \ref{fig:figure5} we would need to rotate each point along the $z$-axis. To calculate the rotation positions we use the Rodrigues' rotation formula as follows
       \begin{center}
       		$\vec{v_{rot}} = \vec{v} \cos\theta+ (\vec{k} \times \vec{v})\sin\theta+\vec{k}(\vec{k}\cdotp \vec{v})(1-\cos\theta)$,
       \end{center}
where $\vec{v}$ is the vector that needs to be rotated, $\vec{k}$ is the axis of rotation and $\theta$ is the angle by which the vector $\vec{v}$ needs to be rotated. Steps for calculating this rotation are listed below:
\begin{enumerate}
\item Calculate the angular separation theta between every branch. This is done by dividing $2\pi$ by the total number of branches.
\item From the origin of rotation, each branch is $\theta$ away from the previous branch. Let $b \in B$ be the current branch and $i_b$ the current index of $b$. Then, $b$ is an angle of $\theta_{rot} = \theta \cdot (i_b-1)$ away from the origin.
\item Rotate $b$ by $\theta_{rot}$ along the $z$-axis relative to the direction between the mUAV and the CH.
\item Following the Rodrigues rotation formula, the $\vec{v}$ which is the formation direction is a vector perpendicular to the direction heading from CH to the mUAV with a magnitude of $\epsilon_d$.
\item  The vector $\vec{v}$, once rotated along the axis of rotation $\vec{k}$, will yield the branch positions along the $z$-axis. Note that at this rotation step, the whole branch will not be in its correct position.
\item The axis of rotation $\vec{k}$ is the direction that is represented by tracing a vector from the CH to the mUAV and normalizing it so that a unit vector is obtained. The values $\vec{v}$ and $\vec{k}$ along with the rotation angle $\theta_{rot}$ for the concerned specific branch is substituted in the Rodrigues rotation formula to yield the positions for every branch.
\item Every branch member $b_i$ occupies an angle of their corresponding $\theta_{rot_i}$ from the origin as shown in Fig \ref{fig:figure5}.
\item With the fractional angle, the actual positions of each branch member can be calculated. The $x$ component is decomposed to be in the branch parent and tracing the fractional angle over the cluster radius to derive the magnitude. The $z$-axis is in the direction of the mUAV. This decomposition is shown in Fig \ref{fig:figure7}.

\end{enumerate}
        
        Let $b$ be a dUAV receiving a rotation message. Let $c$ be the child of $b$.
        Then, $b$ will run Algorithm \ref{algo:rotationPositionsChildren}.
        
        \begin{algorithm}
			\caption{\label{algo:rotationPositionsChildren} Children rotation positions}
			\begin{algorithmic}
            	\Procedure {doRotation}{$\theta_{frac}$, $\vec{d}$, $\vec{v}$, $r_F$, $t$}
            	\State $\alpha_b \gets \theta_{frac} \cdot w_{B_b}$
				
                \State ${x} \gets r_F \cdot \cos(\alpha_b)$
                \State ${z} \gets r_F \cdot \sin(\alpha_b)$
                
                \State $x' \gets \hat{v} \cdot  {x}$
                \State $z' \gets  \hat{d} \cdot  {z}$
            	\State $t_{new} \gets t - \vec{d}$
                \State moveTo($\hat{t}_{new}$)
             	\State send rotation message to $c$ with $(\theta_{frac}, \vec{d},\vec{v},r_F,pos)$
				\EndProcedure
                \State \textbf{end procedure}
            \end{algorithmic}
		\end{algorithm}

\section{Chase Phase}
\label{sec:chase}
During the chase phase, all CHs move towards the mUAV. Because of the continuous execution of the formation algorithm, the resulting movement is also translated to the other members of the clusters. However, the mUAV may be moving towards a certain direction. Thus simply heading directly towards his current position may in some instances not be the shortest path. A strategy to improve the chasing consists in predicting the future position of the mUAV. For this, two positions at different timestamps are compared to each other, forming a movement vector which then can be multiplied by a certain factor in order to obtain the next predicted mUAV position. If needed, multiple points with weight distributions can be used to increase the accuracy of the prediction.
        
        \begin{figure}
			\makebox[\linewidth]{\includegraphics[width=60mm]{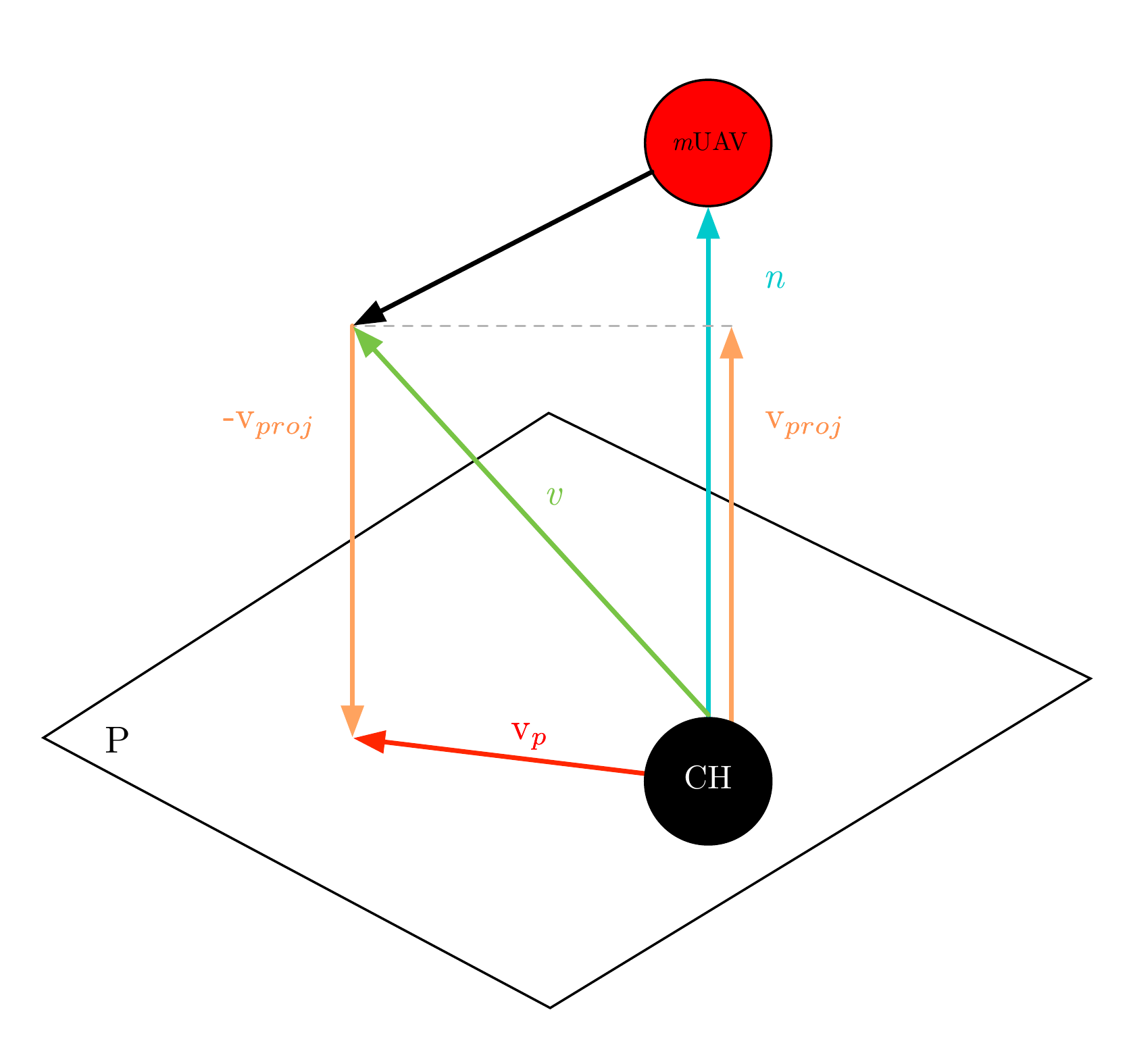}}
			\caption{Predicting the mUAV movement.}
			\label{fig:movementPrediction}
		\end{figure}
        
        As depicted in Fig. \ref{fig:movementPrediction}, the goal is to project the vector $v$ onto the plane $P$, defined by the normal vector $n$. The resulting vector $v_p$ is then added on top of the main heading direction $n$. The complete procedure is shown in Algorithm \ref{algo:movementPreciction}. Note that $\alpha_{x,y}$ represents the angle between $x$ and $y$.
        
        \begin{algorithm}
			\caption{\label{algo:movementPreciction}Chase procedure of CHs including mUAV movement prediction}
			\begin{algorithmic}
				\State $\vec{n} \gets m$UAV$.pos - pos$
                \State $\vec{v} \gets m$UAV$_{predicted} - pos$
                \State $\vec{v}_{proj} \gets \vec{n}.setMag(\cos( \alpha_{\vec{n}, \vec{v}}) \cdot \left\Vert \vec{v} \right\Vert)$
                \State $\vec{v}_p \gets \vec{v} - \vec{v}_{proj}$
                \State $applyForce(\vec{v}_p, w_0)$
                \State $applyForce(\vec{n}, w_1)$
			\end{algorithmic}
		\end{algorithm}		
        
    	Finally, once the distance between the CH and the mUAV is lower than a certain threshold, the enclosure angle of the cluster formation is enlarged, thus trapping the mUAV inside the resulting spherical structure and triggering the escort phase.
        
\section{Escort Phase}
\label{sec:escort}
    The escort phase consists of bringing the previously trapped mUAV outside the flight zone. According to our assumptions, the mUAV will try to avoid any collision with nearby UAVs, and thus is forced to move with them as shown on Fig. \ref{fig:sim-escort}. Optionally, the branches could actively perform anti-escaping blocking maneuvers in order to avoid loosing the mUAV due to larger holes in the formation.
    
    During the process, the CH is in charge of the heading, while its branches maintain their relative positions to the CH. Usually, the shortest path to the flight zone border is taken. If needed, this can be freely adjusted depending on the end goal of the mission.
    
\section{Implementation}
\label{sec:prototype}
    As a proof of concept, a tool that simulates the whole process of the escort mission was developed in JavaScript. This tool contains additional features to model some simplified physics as described hereinafter.
    
     \subsection{Wobbling}
     Every UAV is able to slightly deviate within a given radius from its anchor point. For this, they continuously generate random Perlin Noise\footnote{Perlin Noise: \url{https://mzucker.github.io/html/perlin-noise-math-faq.html}} values for their three movement axes. The resulting pseudo-random movement is supposed to represent the real-world floating instability of UAVs, especially on windy weather conditions. In the case of the mUAV, the wobbling can be used in order to simulate a spontaneous and non-predictable movement, making the chase and escort phases less trivial and thus resulting in a more real-world like scenario.
     
     \subsection{Separation}
     Let $u_{1}$ and $u_2$ be two UAVs. Let $d = \left\Vert u_1.pos - u_2.pos \right\Vert$ be the distance between $u_1$ and $u_2$. Then, if $d < \epsilon_d$, a force vector parallel to $d$ and of amplitude  $\epsilon_d - d$ is applied to $u_{1,2}$, resulting in a separation. When more UAVs are involved, the sum of all produced vectors is applied. If desired, a constant $c$ can be added to each force vector in order to push the UAVs even further apart, making them less likely to stay at the exact borders of the threshold radius. Note that the defense UAVs and the malicious UAV work similarly in terms of collision avoidance principles, that is, they can only move in a given direction if there is no other UAV.
  
  	\subsection{Cumulative force movement logic}
    Multiple forces of different origins may act on some UAV at the same time. For instance, a UAV may at the same time try to head towards a certain direction and actively try to avoid a collision with another UAV. A force can be described as a directional vector $v$ and a weight $w$. At the end of an update cycle, all executed forces are added together, with respect of their weights, to form a cumulative force $v_{sum}$. The amplitude of $v_{sum}$ cannot exceed the maximal velocity of the UAV. The wobbling effect of a UAV is the only movement component that is not translated into force as it is not produced by the UAV itself but by its environment (e.g. wind), implying that the combination of $v_{sum}$ and the produced wobbling movement may exceed the maximal velocity of that UAV.
  
    	\subsection{Visualization}
\begin{figure}

		\begin{minipage}{0.48\linewidth}
			\makebox[\linewidth]{\includegraphics[width=40mm]{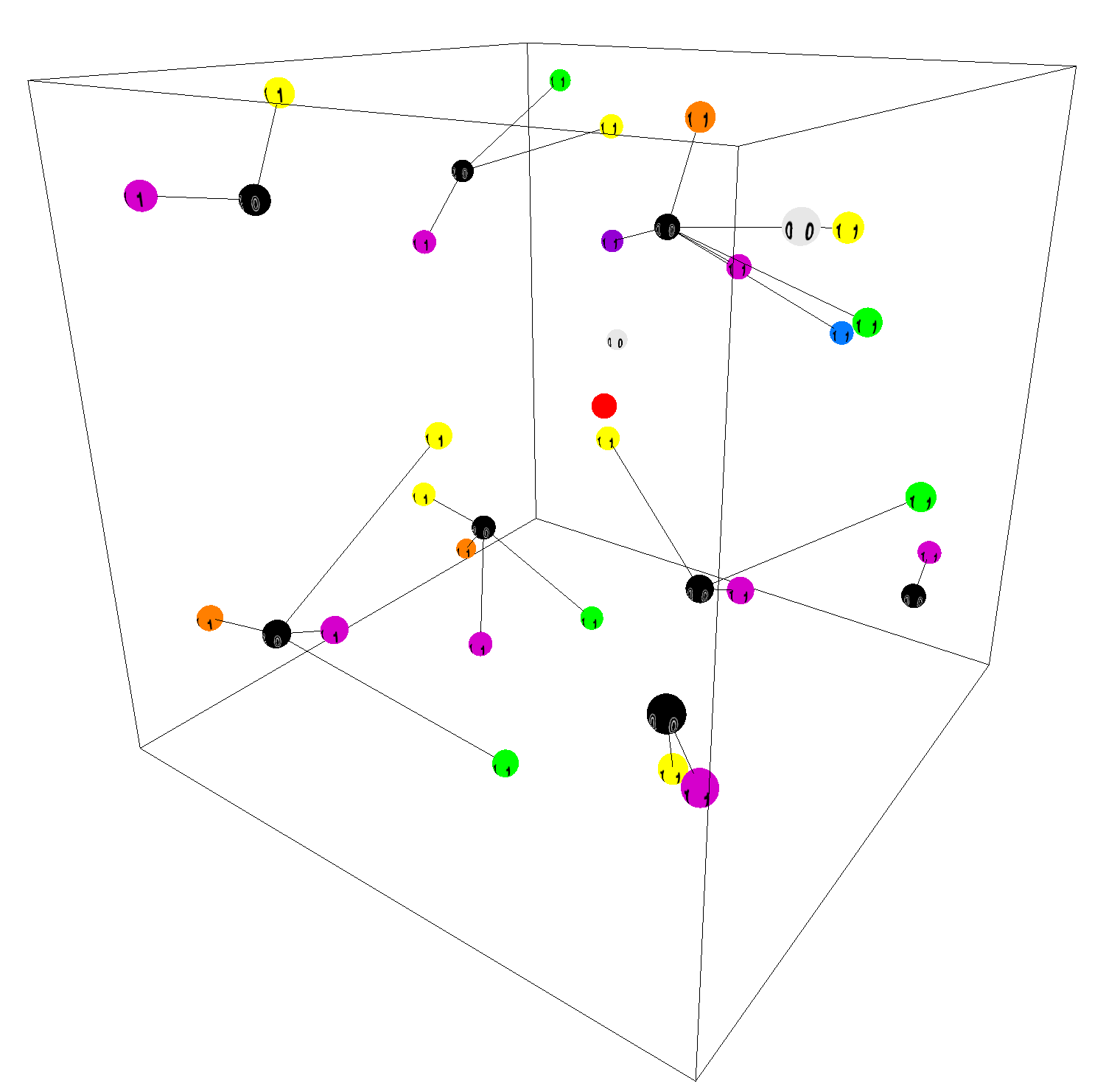}}
			\caption{Clustering.}
			\label{fig:sim-clustering}
		\end{minipage}
        \begin{minipage}{0.49\linewidth}
					\centering
					\includegraphics[width=45mm]{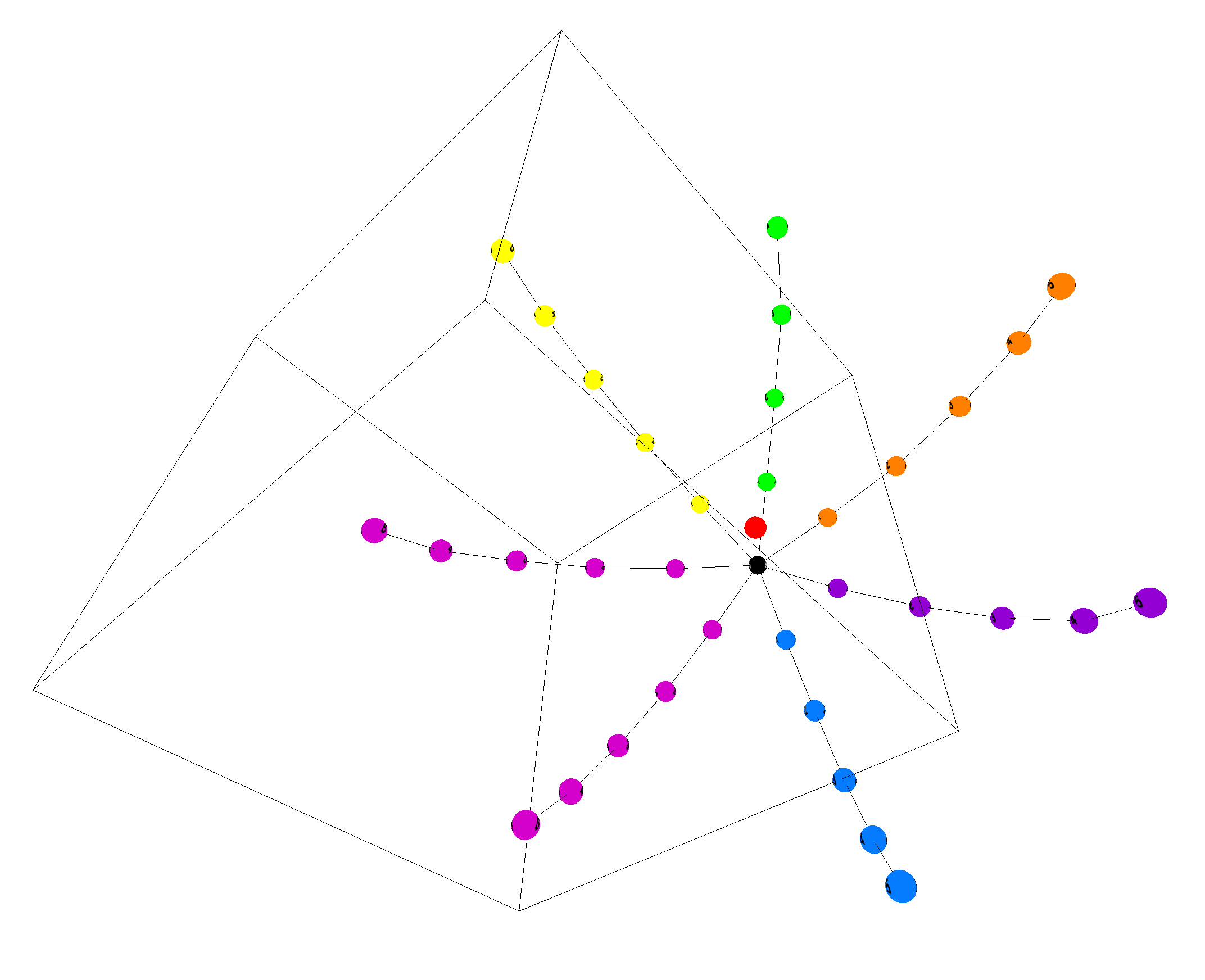}
                    \caption{Formation.}
				\label{fig:sim-formation}
				\end{minipage}
	\end{figure}

	Initially, the simulator creates a set of UAVs randomly distributed within the flight zone and a mUAV. Then each UAV starts running its local clustering algorithms, resulting in cluster formation as illustrated in Fig. \ref{fig:sim-clustering} where black spheres represent CHs and the red sphere represents the mUAV.

\begin{figure}
			\begin{minipage}{0.49\linewidth}
				\makebox[\linewidth]{\includegraphics[width=37mm]{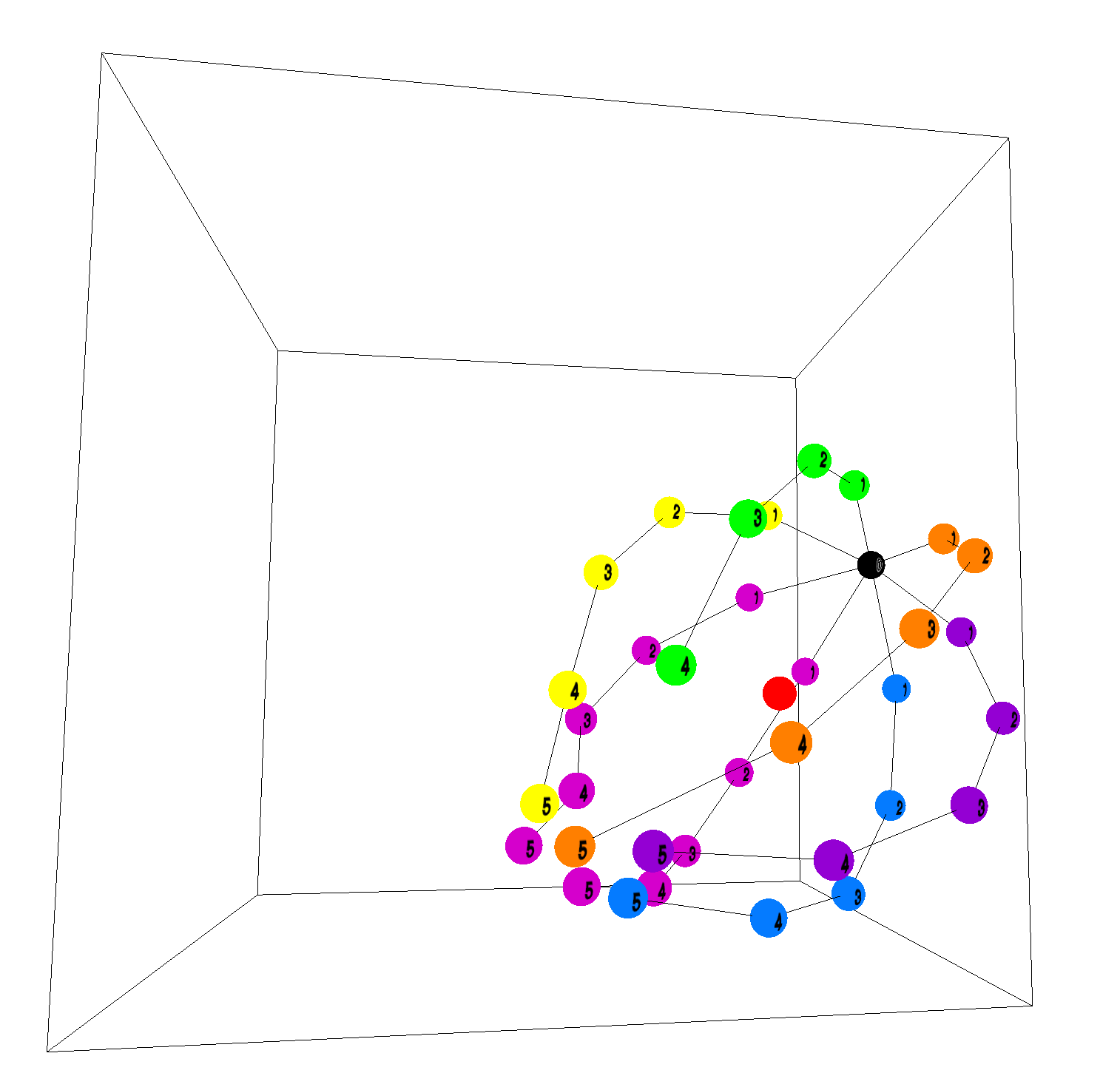}}
				\caption{Chasing.}
				\label{fig:sim-chase1}
			\end{minipage}
			\begin{minipage}{0.49\linewidth}
				\makebox[\linewidth]{\includegraphics[width=49mm]{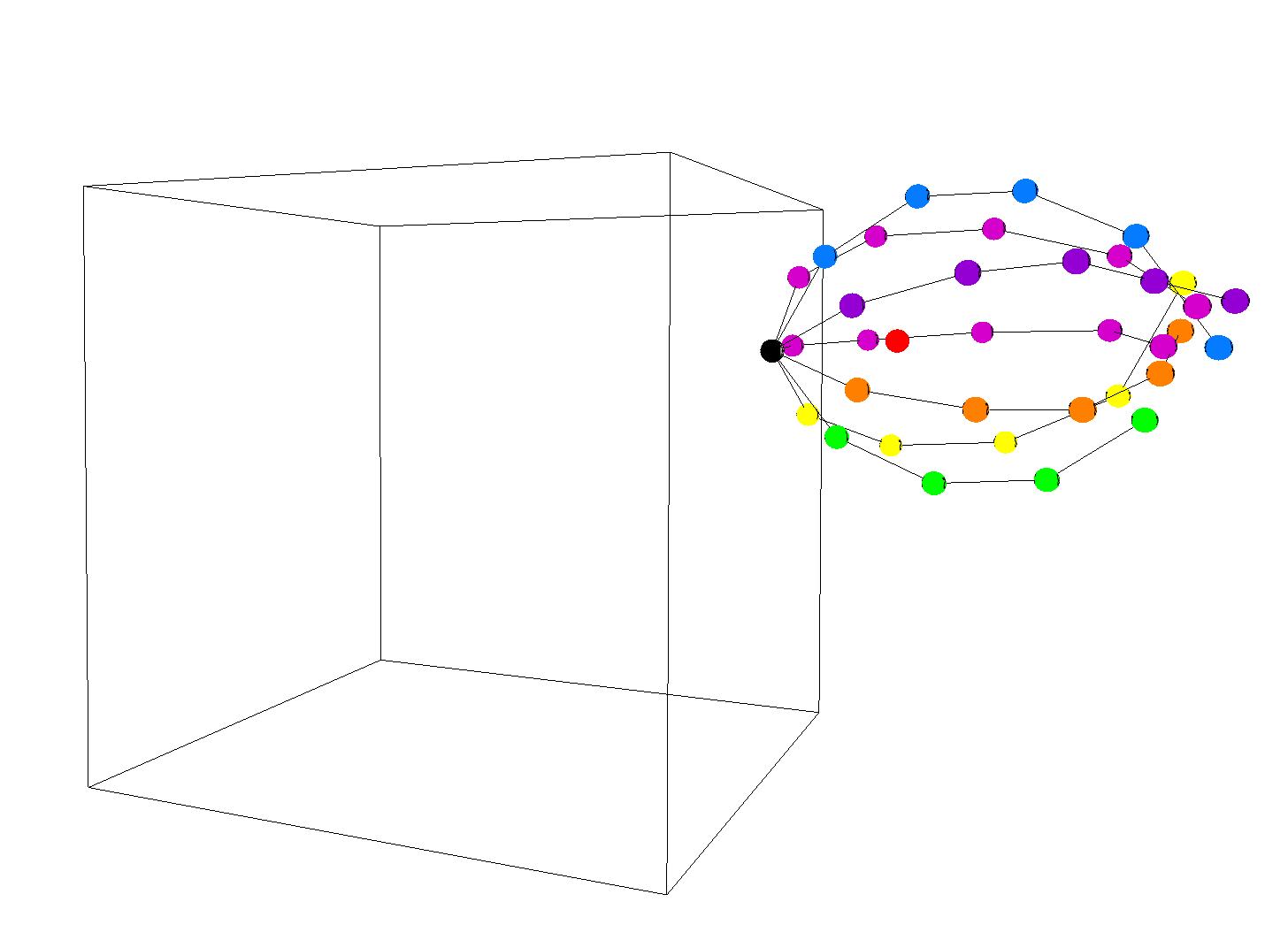}}
			\caption{Escorting.}
			\label{fig:sim-escort}
			\end{minipage}

		\end{figure}

        Fig. \ref{fig:sim-formation} shows a single cluster which is obtained thanks to a merging mechanism between several clusters which is triggered when at least two CH are within their communication range. Fig. \ref{fig:sim-chase1} shows how the formation encircles the mUAV with its different branches, and finally Fig. \ref{fig:escort-mission-progress} presents the mUAV finally escorted outside the flight zone.     	

\section{Simulation Study}
\label{sec:sim-study}
\subsection{Simulation setup and performance metric}

The parameters used for the simulation are presented in Table \ref{tab:sim-params}.

For all simulations, we consider the existent of one mUAV in the center of the flight zone, whereas dUAVs are uniformly positioned at one side of the flight zone. For each simulation, 100 independent simulation runs have been conducted to ensure a statistical significance. 

\begin{table}

\centering
\begin{tabular}{|l|l|}
\hline
Flight zone dimensions & $500\times 500\times 500$ \\ \hline
Number of dUAVs & $20$ \\ \hline
Communication Range & $100$ \\ \hline
dUAV Wobbling Radius & $50$ \\ \hline
mUAV Wobbling Radius & $150$ \\ \hline
dUAV Collision Threshold & $40$ \\ \hline
mUAV Collision Threshold & $60$ \\ \hline
Number of Branches & $3$ \\ \hline
Angular separation $\theta$ & $\pi/2$ \\ \hline
UAV speed & $0.8$ \\ \hline
UAV radius & $10$ \\ \hline
\end{tabular}
\caption{Simulation parameters.}
\label{tab:sim-params}
\end{table}

The performance of the experiments is measured by the time of the dUAVs needed to successfully escort the mUAV outside the flight zone.

\subsection{Experiments and results}

By performing the Anderson-Darling normality test over 100 simulations, we obtained a \emph{p}-value of $0.247$, thus giving strong evidence that the data follows a normal distribution.

\begin{figure}
  \centering
  \includegraphics[width=80mm]{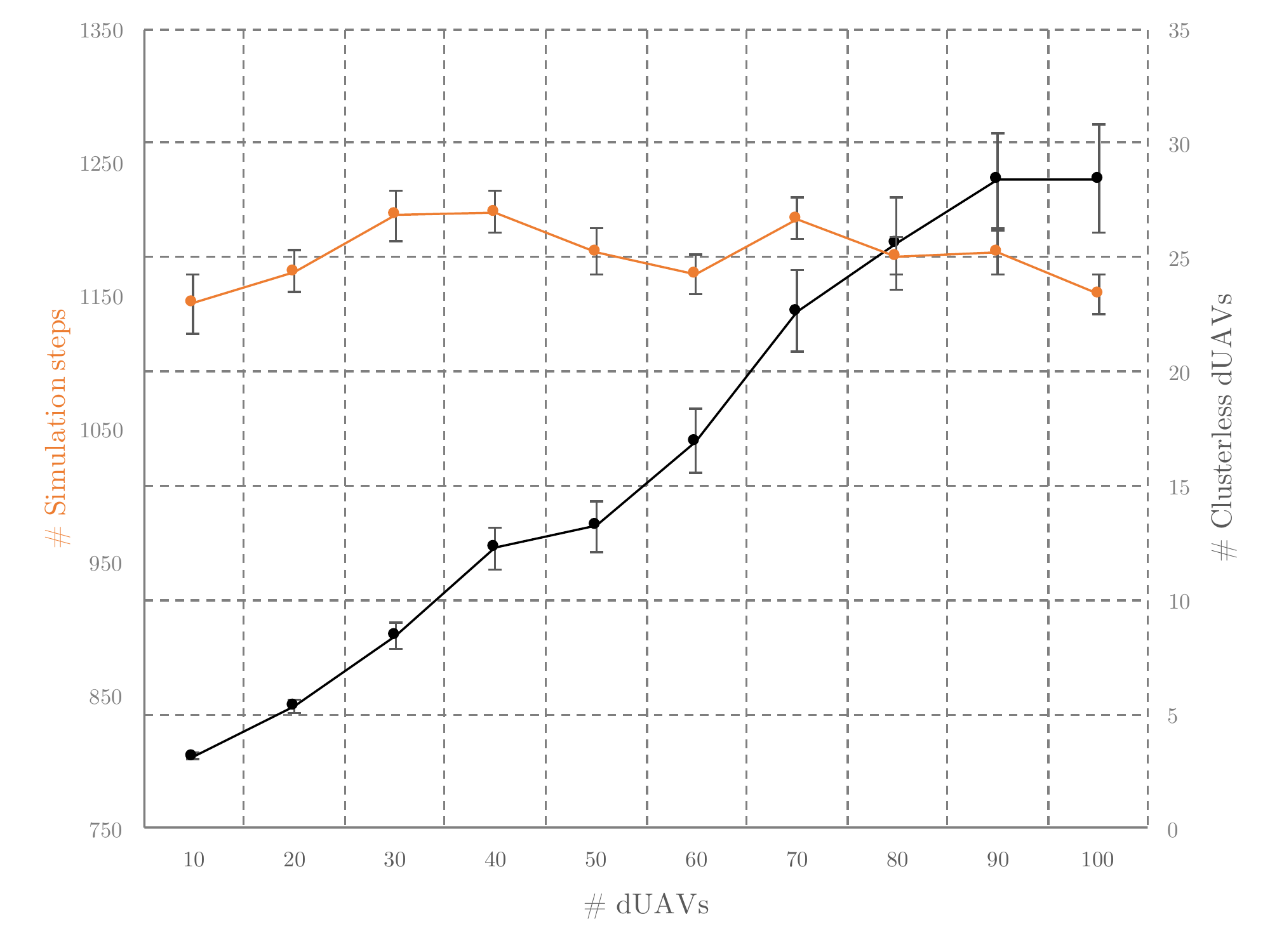}
  \caption{Dependency between number of dUAVs and simulation times.}
  \label{fig:output0}
\end{figure}

In Fig. \ref{fig:output0}, we can observe that the impact of the number of dUAVs on the escort time is less restricted. However, the number of clusterless dUAVs, i.e. dUAVs not affiliated to any cluster, increases proportionally to the total amount of dUAVs in the flight zone. Conclusively, since these UAVs do not contribute to the escort mission, resources can be saved here by deploying fewer dUAVs.

\begin{figure}
  \centering
  \includegraphics[width=80mm]{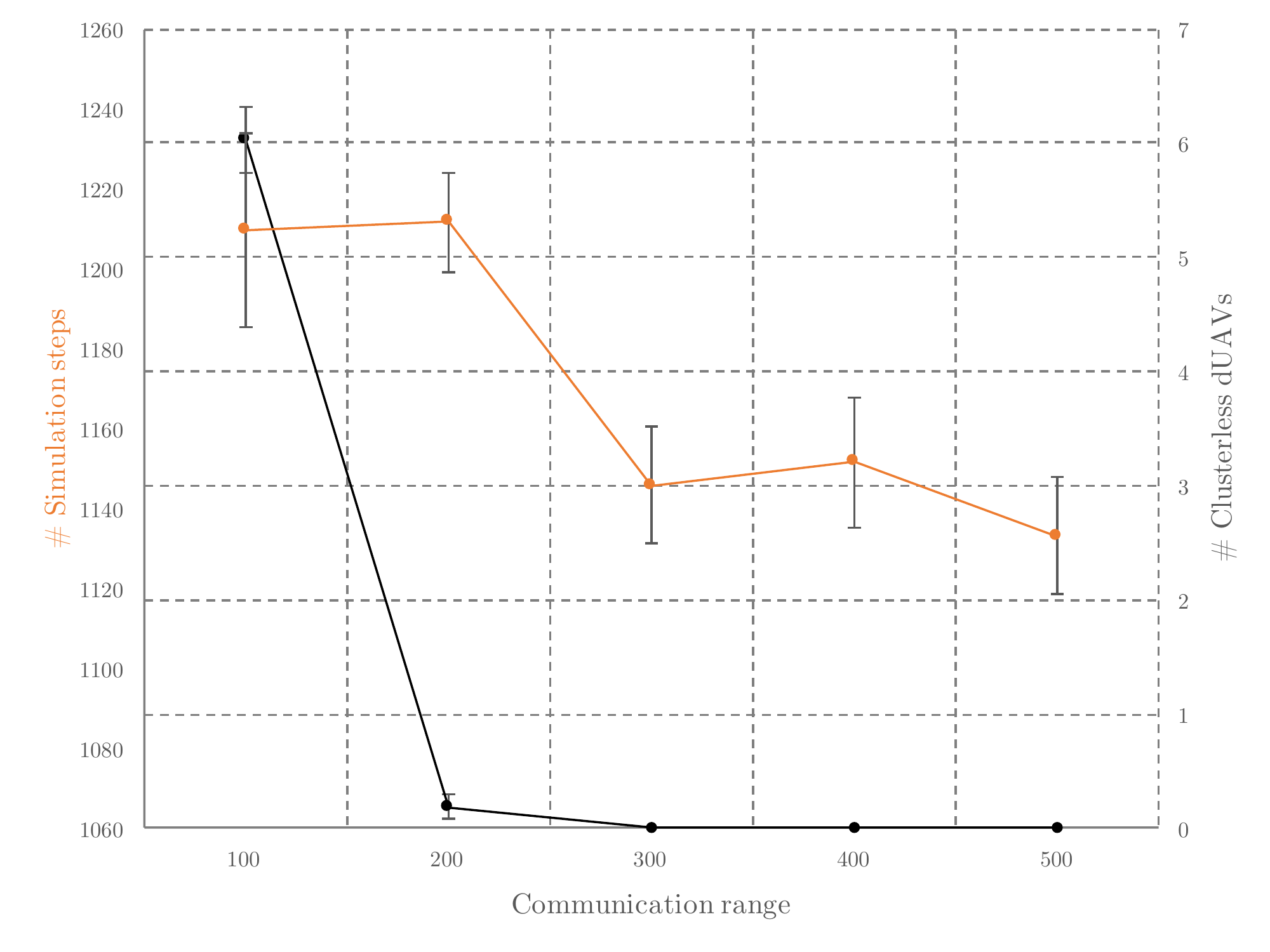}
  \caption{Dependency between communication range and simulation times.}
  \label{fig:output1}
\end{figure}

The performance of the simulation improves noticeably when choosing a larger communication range between dUAVs until reaching a point of diminishing returns as depicted in Fig \ref{fig:output1}. Intuitively, the number of clusterless dUAVs decreases with larger communication range as each UAV can locate its neighbors.

\begin{figure}
  \centering
  \includegraphics[width=80mm]{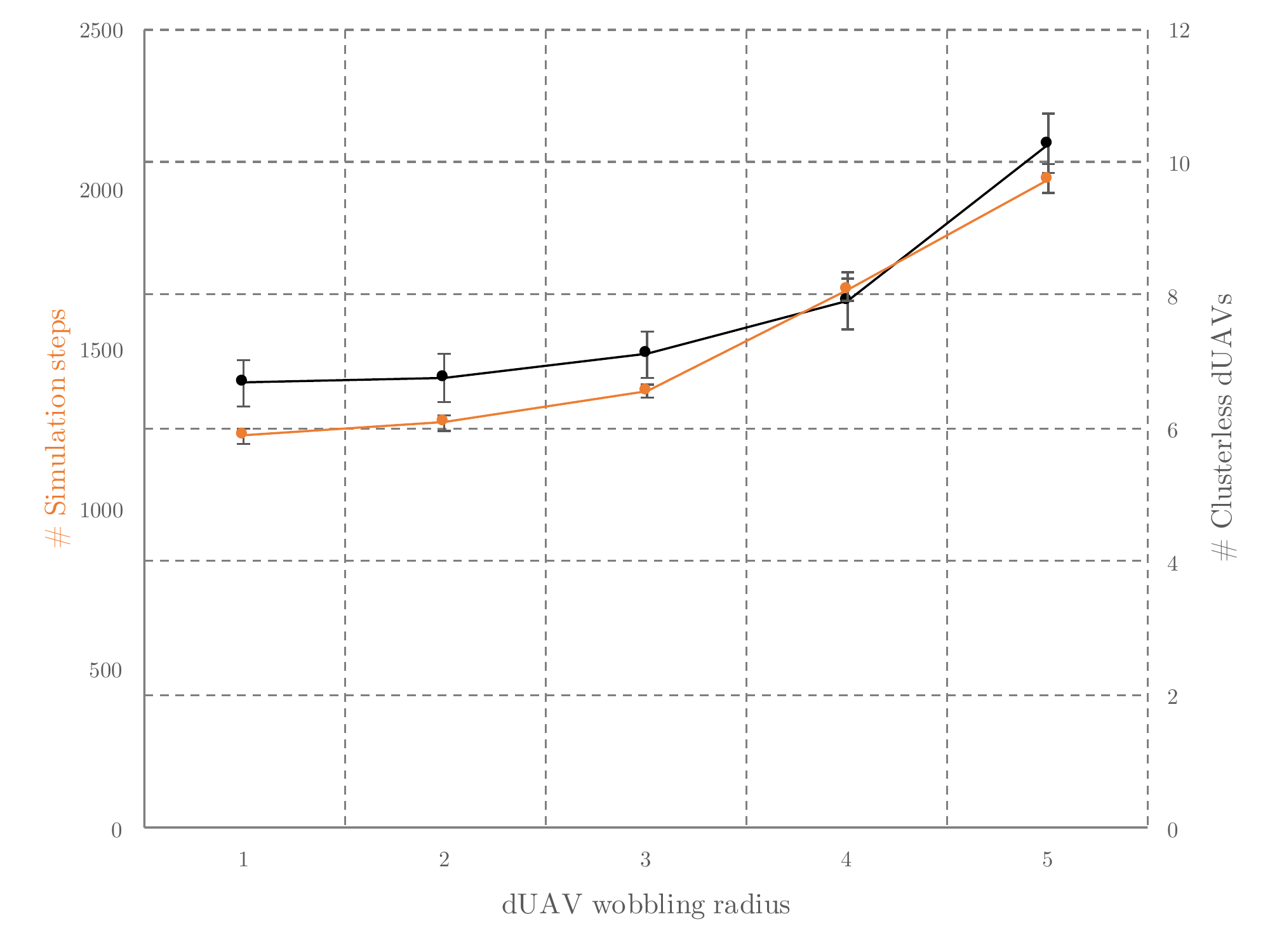}
  \caption{Dependency between dUAV wobbling radius and simulation times.}
  \label{fig:output2}
\end{figure}

From Fig. \ref{fig:output2}, it is clear that the wobbling of UAVs has a negative impact on the escort times and therefore should be minimized as much as possible.

\begin{figure}
  \centering
  \includegraphics[width=80mm]{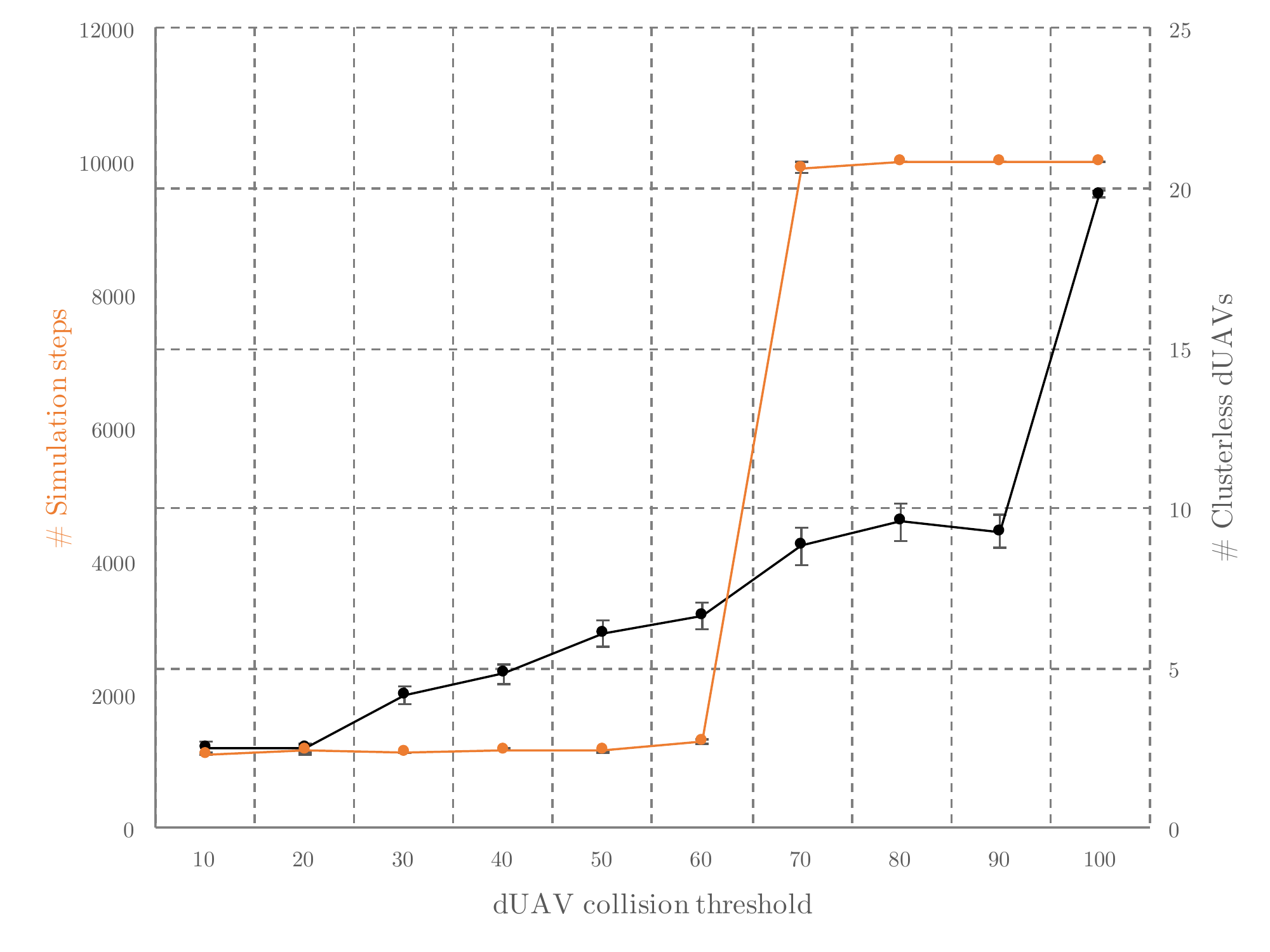}
  \caption{Dependency between dUAV collision threshold and simulation times.}
  \label{fig:output4}
\end{figure}

\begin{figure}
  \centering
  \includegraphics[width=80mm]{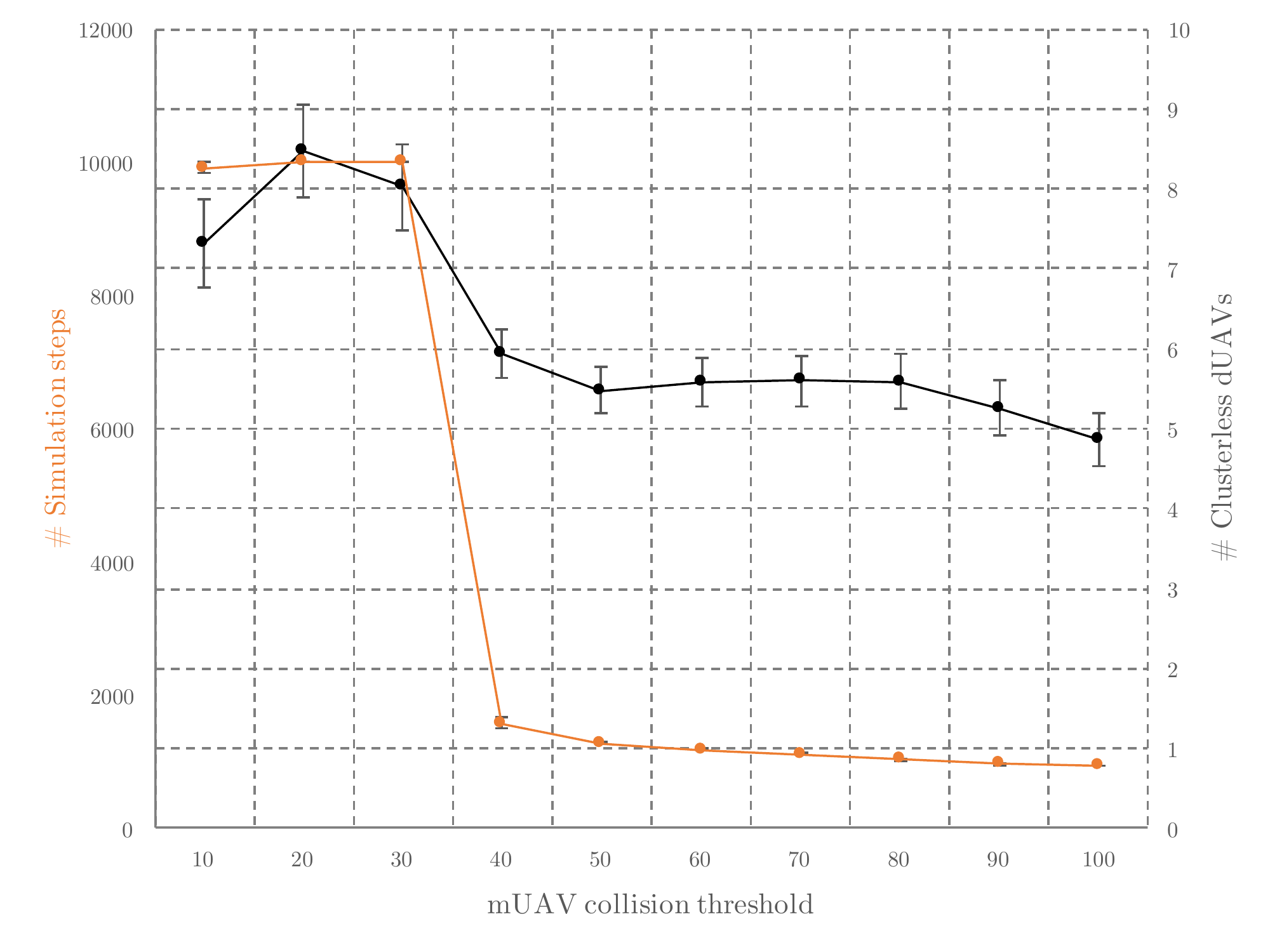}
  \caption{Dependency between mUAV collision threshold and simulation times.}
  \label{fig:output5}
\end{figure}

Fig. \ref{fig:output4} shows that the simulation fails when choosing a dUAV collision threshold over 60 as well as when the mUAV collision threshold is below 40 (Fig. \ref{fig:output5}). When looking at the initial configuration, we see that the collision threshold for dUAV and mUAV has been chosen to be at 40 and 60 respectively. In fact, as soon as the value of the mUAV surpasses the one of the dUAVs, the simulation fails since the dUAVs are unable to push the mUAV outside its current position.

\begin{figure}
  \centering
  \includegraphics[width=80mm]{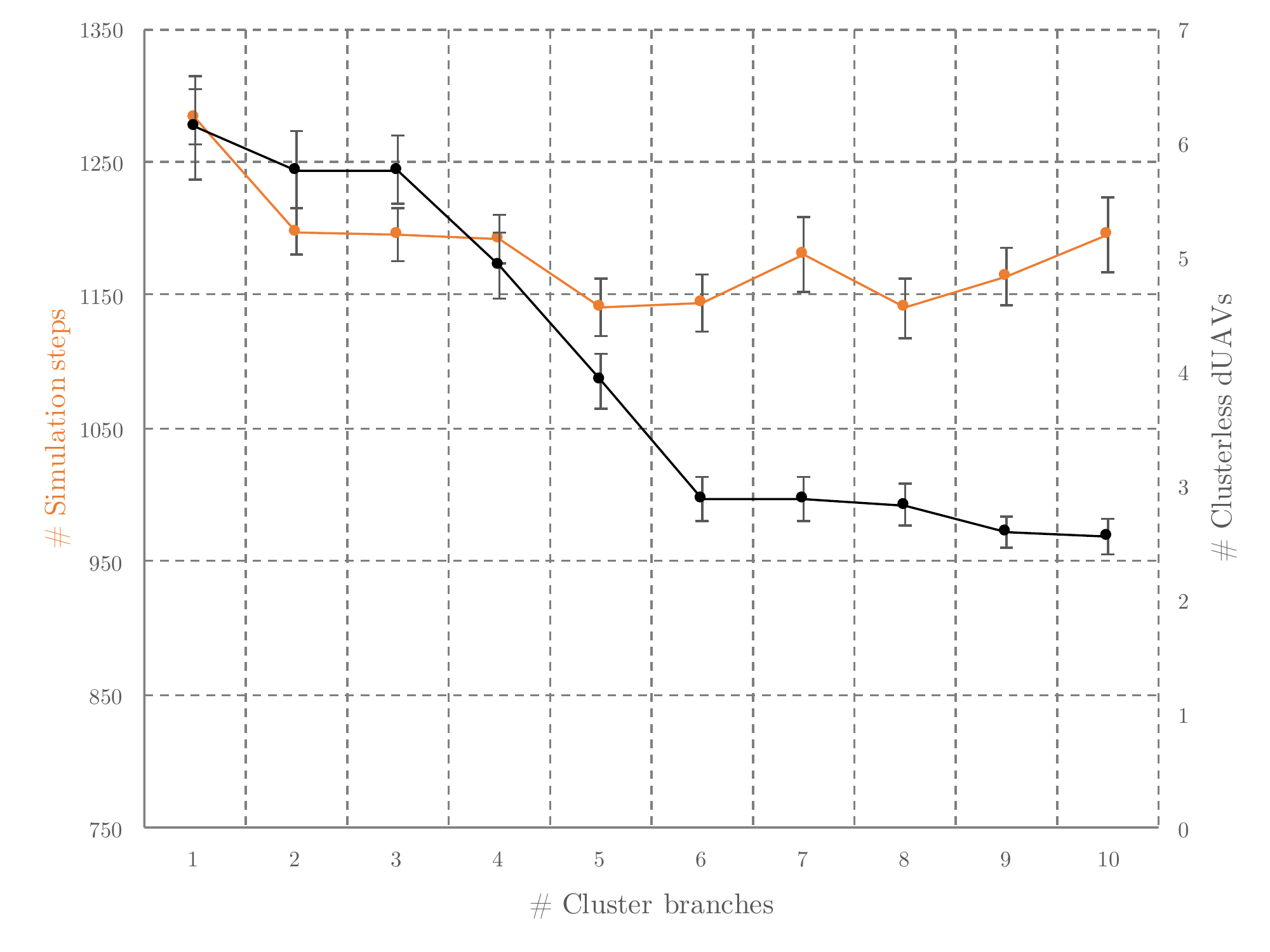}
  \caption{Dependency between number of branches and simulation times.}
  \label{fig:output6}
\end{figure}
Finally, Fig. \ref{fig:output5} shows that starting from 2, the number of branches does not have a significant impact on the simulation times for the specified initial configuration.

\section{Conclusions and Future Work} 
\label{sec:conclusion}
A comprehensive UAV defense system has been proposed, which is able to deploy auto-organized defense UAVs (dUAV) and create a intercept- and capture-formation to escort malicious UAVs (mUAV) outside a flight zone. 

The most outstanding features and contributions of the presented approach are the balanced clustering to realize the intercept- and capture-formation. Additionally, the approach consists of a modular design containing the phases deployment, clustering, formation, chase, escort. All parts of the approach are fully localized, and the resulting networked defense UAV swarm is resilient against communication losses.

\section*{Acknowledgment}
This work is partially funded by the ILNAS (Institut Luxembourgeois de la Normalisation, de l'Accr\'editation, de la S\'ecurit\'e et qualit\'e des produits et services) under the research project ``Digital Trust for Smart ICT''.

\balance  
\bibliographystyle{IEEEtranS}

\begin{thebibliography}{10}
\providecommand{\url}[1]{#1}
\csname url@samestyle\endcsname
\providecommand{\newblock}{\relax}
\providecommand{\bibinfo}[2]{#2}
\providecommand{\BIBentrySTDinterwordspacing}{\spaceskip=0pt\relax}
\providecommand{\BIBentryALTinterwordstretchfactor}{4}
\providecommand{\BIBentryALTinterwordspacing}{\spaceskip=\fontdimen2\font plus
\BIBentryALTinterwordstretchfactor\fontdimen3\font minus
  \fontdimen4\font\relax}
\providecommand{\BIBforeignlanguage}[2]{{%
\expandafter\ifx\csname l@#1\endcsname\relax
\typeout{** WARNING: IEEEtranS.bst: No hyphenation pattern has been}%
\typeout{** loaded for the language `#1'. Using the pattern for}%
\typeout{** the default language instead.}%
\else
\language=\csname l@#1\endcsname
\fi
#2}}
\providecommand{\BIBdecl}{\relax}
\BIBdecl

\bibitem{Brust2016VBCAAV}
M.~R. Brust, M.~I. Akbas, and D.~Turgut, ``Vbca: A virtual forces clustering
  algorithm for autonomous aerial drone systems,'' in \emph{IEEE SysCon}, 2016.

\bibitem{brust2007adaptive}
M.~R. Brust, H.~Frey, and S.~Rothkugel, ``Adaptive multi-hop clustering in
  mobile networks,'' in \emph{Proceedings of the 4th international conference
  on mobile technology, applications, and systems and the 1st international
  symposium on Computer human interaction in mobile technology}.\hskip 1em plus
  0.5em minus 0.4em\relax ACM, 2007, pp. 132--138.

\bibitem{brust2008dynamic}
------, ``Dynamic multi-hop clustering for mobile hybrid wireless networks,''
  in \emph{Proceedings of the 2nd international conference on Ubiquitous
  information management and communication}.\hskip 1em plus 0.5em minus
  0.4em\relax ACM, 2008, pp. 130--135.

\bibitem{brust2010lswtc}
M.~R. Brust, C.~H. Ribeiro, D.~Turgut, and S.~Rothkugel, ``Lswtc: A local
  small-world topology control algorithm for backbone-assisted mobile ad hoc
  networks,'' in \emph{Local Computer Networks (LCN), 2010 IEEE 35th Conference
  on}.\hskip 1em plus 0.5em minus 0.4em\relax IEEE, 2010, pp. 144--151.

\bibitem{brust2015networked}
M.~R. Brust and B.~M. Strimbu, ``A networked swarm model for uav deployment in
  the assessment of forest environments,'' in \emph{Intelligent Sensors, Sensor
  Networks and Information Processing (ISSNIP), 2015 IEEE Tenth International
  Conference on}.\hskip 1em plus 0.5em minus 0.4em\relax IEEE, 2015, pp. 1--6.

\bibitem{Gregoire:2015:MultiLevel}
G.~Danoy, M.~R. Brust, and P.~Bouvry, ``Connectivity stability in autonomous
  multi-level {UAV} swarms for wide area monitoring,'' in \emph{Proceedings of
  the Fifth ACM International Symposium on Development and Analysis of
  Intelligent Vehicular Networks and Applications}, ser. DIVANet '15.\hskip 1em
  plus 0.5em minus 0.4em\relax ACM, 2015.

\bibitem{humphreys2015statement}
T.~Humphreys, ``Statement on the security threat posed by unmanned aerial
  systems and possible countermeasures,'' \emph{Oversight and Management
  Efficiency Subcommittee, Homeland Security Committee, Washington, DC, US
  House}, 2015.

\bibitem{kerns2014unmanned}
A.~J. Kerns, D.~P. Shepard, J.~A. Bhatti, and T.~E. Humphreys, ``Unmanned
  aircraft capture and control via gps spoofing,'' \emph{Journal of Field
  Robotics}, vol.~31, no.~4, pp. 617--636, 2014.

\bibitem{maxa2017survey}
J.-A. Maxa, M.-S.~B. Mahmoud, and N.~Larrieu, ``Survey on uaanet routing
  protocols and network security challenges,'' \emph{Ad Hoc \& Sensor Wireless
  Networks}, 2017.

\bibitem{mitchell2014adaptive}
R.~Mitchell and R.~Chen, ``Adaptive intrusion detection of malicious unmanned
  air vehicles using behavior rule specifications,'' \emph{IEEE Transactions on
  Systems, Man, and Cybernetics: Systems}, vol.~44, no.~5, pp. 593--604, 2014.

\bibitem{10945/5700}
M.~F. Munoz, ``Agent-based simulation and analysis of a defensive uav swarm
  against an enemy {UAV} swarm,'' Master's thesis, Naval Postgraduate School,
  Monterey, USA, 2011.

\bibitem{olfati2006flocking}
R.~Olfati-Saber, ``Flocking for multi-agent dynamic systems: Algorithms and
  theory,'' \emph{IEEE Transactions on automatic control}, vol.~51, no.~3, pp.
  401--420, 2006.

\bibitem{reynolds1987flocks}
C.~W. Reynolds, ``Flocks, herds and schools: A distributed behavioral model,''
  \emph{ACM SIGGRAPH computer graphics}, vol.~21, no.~4, pp. 25--34, 1987.

\bibitem{Eagles}
\BIBentryALTinterwordspacing
J.~J. Roberts, \emph{France Is Training Eagles to Kill Drones}, 2017 (accessed
  August 14, 2017). [Online]. Available:
  \url{http://fortune.com/2017/02/22/drones-eagles-france}
\BIBentrySTDinterwordspacing

\bibitem{10945/17462}
\BIBentryALTinterwordspacing
U.~Soylu, ``Multi-target tracking for swarm vs. swarm uav systems,'' Master's
  thesis, Naval Postgraduate School, Monterey, USA, 2012. [Online]. Available:
  \url{https://calhoun.nps.edu/handle/10945/17462}
\BIBentrySTDinterwordspacing

\bibitem{wesson2013hacking}
K.~Wesson and T.~Humphreys, ``Hacking drones.'' \emph{Scientific American},
  vol. 309, no.~5, p.~54, 2013.

\end{thebibliography}

\end{document}